\documentclass[11pt]{article}
\usepackage{geometry}             
\usepackage{graphicx}
\usepackage{amssymb}
\usepackage{amsmath}
\usepackage{epstopdf}
\usepackage{comment}
\usepackage{cite}
\usepackage{color}

\definecolor{brightpink}{rgb}{1.0, 0.0, 0.5}
\usepackage[hyperindex=true,
          pdfstartview=FitH,
          bookmarksnumbered=true,
          bookmarksopen=true,
          citecolor=blue,
          linkcolor=blue,
          colorlinks=true,
          unicode]{hyperref}

\newcommand{\be}{\begin{equation}}
\newcommand{\ee}{\end{equation}}
\newcommand{\bea}{\begin{eqnarray}}
\newcommand{\eea}{\end{eqnarray}}

\begin{document}
	
	\title{Gauss-Bonnet modification to Hawking evaporation of AdS black holes in massive gravity}
	\author{
		Hao Xu
		$$\thanks{{\em Corresponding author, E-mail }:\href{mailto: haoxu@yzu.edu.cn }
		{ haoxu@yzu.edu.cn}}  , Yun Du$${}
		\vspace{5pt}\\
		\small  $$Center for Gravitation and Cosmology, College of Physical Science and Technology,\\
		\small Yangzhou University, Yangzhou, 225009, China\\
	}

	\date{}
	\maketitle

\begin{abstract}
{
The Stefan-Boltzmann law can estimate particle emission power and lifetime of a black hole. However, in modified gravity theories, new parameters in the action can cause qualitative changes in thermodynamic quantities, thus obtaining specific thermodynamic properties often requires complicated calculation with higher degree equations. In this work, we aim to provide a general model-independent description of the evolution of AdS black holes, using Gauss-Bonnet massive gravity as an example. We prove that the impact factor of an infinitely large AdS black hole is equal to the effective AdS radius, and the black hole is able to evaporate an infinite amount of mass in finite time, so that the lifetime of the black hole depends on the final state temperature in the evaporation process. The black hole will evaporate out when the final state temperature diverges and will transform into a remnant when the temperature is zero. Since we have analyzed massive gravity in detail, we can introduce the Gauss-Bonnet term to study how it affects the thermodynamic quantities. We obtain the final states for different parameter intervals, study the qualitative properties of black hole evaporation, and classify the associated cases. This method can also be applied to other models.
}

\end{abstract}

\section{Introduction}

General relativity may not be the final word on the gravitational field for at least two reasons. First, despite the success of general relativity in describing physics on the scale of the solar system, there are still a number of unanswered questions when it is applied to larger scales, such as its inconsistency with the observed rotation curves of galaxies and the accelerating expansion of the cosmos. The second, and perhaps the more compelling reason, is that general relativity is a classical theory, whereas the world is fundamentally quantum mechanical. The search for a working theory of quantum gravity is the driving force behind a great deal of research in theoretical physics today, and much has been learned along the way, but convincing success remains out of reach.

To resolve the observational inconsistency at the cosmological scale, ``dark matter'' and ``dark energy'' have been introduced to explain these anomalies. However, despite great efforts to find candidates for them, their true identities remain unknown. So we might ask whether we can modify gravity theory to explain the physics on these larger scales, while preserving the known behavior on solar scales. Put another way, any viable modified gravity theory of dark matter and dark energy would have to be reducible to general relativity, which can be achieved by adding extra terms to the original Einstein-Hilbert action.

One of the candidates for modified gravity is massive gravity. It is an extension of general relativity by giving the graviton a non-zero mass. The first attempt to derive such a theory was made by Fierz and Pauli in 1939 \cite{Fierz:1939ix}, and massive gravity has made great progress since then \cite{VanNieuwenhuizen:1973fi,vanDam:1970vg,Zakharov:1970cc,Vainshtein:1972sx,Boulware:1973my}. After solving a number of problems, in particular by introducing nonlinear terms to exorcise the Boulware-Deser ghost, we now have a ghost-free theory known as dRGT massive gravity \cite{deRham:2010ik,deRham:2010kj,Hassan,HassanI,HassanII}, which in the rest of this paper we will refer to as massive gravity. Although massive gravity suffers from other problems, such as causality violation \cite{1306.5457,1410.2289} and the lack of a stable FLRW solution when the background is chosen to be Minkowski \cite{DAmico, 1304.0484}, the theoretical and phenomenological advantages of massive gravity have stimulated much research in the literature.

On the other hand, from a modern perspective, general relativity may be just an effective field theory in the low energy limit of an unknown fundamental quantum theory. The exact form of this theory is not yet clear to us, but there have been some speculations. Inspired by heterotic string theory and six-dimensional Calabi-Yau compactifications of M-theory, the Gauss-Bonnet squared terms have been introduced into the Einstein-Hilbert action to analyze the properties of effective gravity theory \cite{Boulware:1985wk,Callan:1988hs}. The resulting field equations contain no more than second derivatives of the metric and have been shown to be free of ghosts. Studies of Gauss-Bonnet gravity and its generalizations have attracted much attention \cite{Cai:2001dz,Cvetic:2001bk,Torii:2005xu,Konoplya:2010vz,Xu:2013zea,Xu:2014tja,Xu:2015hba,Sun:2016til,Konoplya:2017ymp,Konoplya:2017zwo}, but this theory exists only in higher spacetime dimensions because the Gauss-Bonnet term is topologically invariant in four dimensions, so it cannot give non-trivial gravitational dynamics. Fortunately, In \cite{Glavan:2019inb}, Glavan and Lin try to extend this theory to four dimensions by rescaling the coupling constant of the Gauss-Bonnet term by a factor of $1/(d - 4)$ and taking the limit $d\rightarrow 4$. Remarkably, from their point of view, this theory can avoid Lovelock's theorem and is free of Ostrogradsky instability. This treatment in  has led to much discussion, see e.g. \cite{Lu:2020iav,Guo:2020zmf,Kumar:2020owy,Zhang:2020qam,Arrechea:2020evj,Yang:2020jno,Konoplya:2020cbv,Zhang:2020sjh,Mahapatra:2020rds,Gurses:2020ofy,Arrechea:2020gjw,Gurses:2020rxb,Shu:2020cjw}.

As mentioned above, the massive gravity can be used to explain dark matter and dark energy, while the Gauss-Bonnet term is used to study effective gravity theory, so it is natural to think of combining them. In this work we would like to study the effect of the Gauss-Bonnet term on the evaporation of the black hole in massive gravity. Hawking radiation allows black holes to emit particles, leaving us with the controversial information loss paradox. However, we can always calculate the particle emission power and estimate the lifetime of the black hole by using Stefan-Boltzmann law. The lifetime of the black hole depends on the gravity theory and the boundary conditions. For example, for a static spherically symmetric black hole in four-dimensional asymptotically flat spacetime in Einstein gravity, its lifetime obeys the relation $t\sim M_0^3$ associated with the initial black hole mass $M_0$ \cite{Page:1976df}. {Since current observations indicate that the universe may be de Sitter, it seems natural that we should study evaporation in de Sitter spacetime. However, there are still some problems with black hole thermodynamics and quantum field theory in de Sitter spacetime that are not yet understood. For example, it is hard to define the thermal equilibrium state due to the varying temperatures of the black hole horizon and the cosmological horizon, and there is no well-defined asymptotic infinity to normalize the Killing vector field.}

{On the other hand, AdS spacetime does not suffer from the above problems. Since AdS/CFT has also been recognized as a bridge to further understanding of quantum gravity \cite{Maldacena:1997re,Witten:1998qj}, studying black hole evaporation in AdS background might help us understand real-world physics}. By imposing an absorbing AdS boundary condition and considering only massless emitted particles, Page found that the lifetime of a four-dimensional static spherically symmetric AdS black hole in Einstein gravity is finite for any infinitely large initial mass and is of the order of $\ell^3$, where $\ell$ is the AdS radius \cite{Page:2015rxa}\footnote{{We can not discuss infinitely large black hole in de Sitter spacetime. Given the relevant parameters, there is an upper limit to the mass of black holes in de Sitter spacetime, so increasing the radius of a black hole causes its radius to converge to the cosmological horizon, eventually reaching the Nariai limit \cite{Nariai1999,Ginsparg1983}}}. The study of black hole evaporation has been extended to the charged case \cite{Hiscock:1990ex,Xu:2019wak}, reflected boundary condition \cite{Ling:2021nad} and many different theories of modified gravity \cite{Xu:2019krv,Xu:2018liy,Xu:2020xsl,Wu:2021zyl}, including the massive gravity \cite{Hou:2020yni}.

For the AdS black hole in massive gravity, it can either evaporate completely or become a remnant at late time, depending on the specific parameters that determine the temperature of the final state. However, if we add the Gauss-Bonnet term, the situation becomes more complicated. {The Gauss-Bonnet massive gravity is a combination of two more well-known models, each of which is easy to solve individually and more difficult when taken together. The two most important physical quantities in the black hole evaporation process, cross section and temperature, become difficult to analyze due to the increase of degree in equations, so we can no longer calculate and classify all cases analytically as in the case of massive gravity \cite{Hou:2020yni}. This situation will likely become more critical in other complex models.}

{Therefore, in this work we expect to give a general model-independent description of the evolution of AdS black holes, using Gauss-Bonnet massive gravity as an example. We look forward to giving some conclusions and methods with a view to being able to deal with a class of models rather than a single model. We will consider the lifetime of an infinitely large AdS black hole in two parts: first, whether it can evaporate an infinite amount of mass in finite time, and second, whether the temperature of the final state is likely to be zero. For the first question, we will prove that \emph{an infinitely large AdS black hole is indeed capable of evaporating an infinite amount of mass in finite time}, so that we do not have to worry about infinite mass and can therefore focus on the zero point of the temperature. For the second question, since we already have a detailed analysis of massive gravity, we can introduce Gauss-Bonnet parameter to see how it corrects the thermodynamic quantities and perform a qualitative analysis. We can obtain the final states for different parameter intervals, study the qualitative properties of black hole evaporation, and classify the corresponding cases. In other models, if we can find one that has been rigorously solved, we can use it as a basis for investigating other parameter corrections.}

The remainder of the paper is organized as follows. In section \ref{section2} we give a brief review of the AdS black hole solution and thermodynamics in four-dimensional Gauss-Bonnet massive gravity. In section \ref{section3} we investigate the cross section and emission rate of an infinitely large AdS black hole, and we can find the impact factor of the black hole corresponds to the effective AdS radius and the integral of the Stefan-Boltzmann law is convergent at infinity, meaning the infinitely large AdS black hole can evaporate infinite mass in finite time. In section \ref{section4} we use the Gauss-Bonnet term as a correction term to classify the final states for different parameter intervals. In the last section we summarize the result. We set the speed of light in vacuum, the gravitational constant, the Planck constant and the Boltzmann constant all equal to unity.

\section{AdS Black Hole solution and thermodynamics in four-dimensional Gauss-Bonnet massive gravity}
\label{section2}

In this section we give a brief review of the AdS black hole solution and its thermodynamics in Gauss-Bonnet massive gravity. Detailed result can be found in \cite{Upadhyay:2022axg}. The action of Gauss-Bonnet massive gravity gravity can be written as Hilbert-Einstein action with the Gauss-Bonnet term and a suitable nonlinear interaction terms given by
\begin{equation}
{I}=\frac{1}{16\pi }\int \text{d}^{4}x\sqrt{-g}\left[ R-2\Lambda + {\alpha}  {\mathcal L_{GB}} +m^{2}\sum_{i} c_i\mathcal{U}_i%
(g,f)\right] ,  \label{action}
\end{equation}%
where $R$ is the Ricci scalar, $\Lambda$ is the negative cosmological constant, $\alpha$ is the Gauss-Bonnet coupling coefficient with dimension $(\text{length})^2$, $\mathcal{L}_{\text{GB}}=\mathcal{R}_{\mu\nu\gamma\delta}
\mathcal{R}^{\mu\nu\gamma\delta}-4\mathcal{R}_{\mu\nu}\mathcal{R}^{\mu\nu}+\mathcal{R}^{2}$ is the Gauss-Bonnet term, and $\mathcal{U}$ is the effective potential of graviton with nonzero graviton mass $m$ and coefficients $c_i$. The effective potential $\mathcal{U}$ reads
\begin{eqnarray}
\mathcal{U}_{1} &=&\left[ \mathcal{K}\right],  \notag \\
\mathcal{U}_{2} &=&\left[ \mathcal{K}\right] ^{2}-\left[ \mathcal{K}^{2}%
\right] ,  \notag \\
\mathcal{U}_{3} &=&\left[ \mathcal{K}\right] ^{3}-3\left[ \mathcal{K}\right] %
\left[ \mathcal{K}^{2}\right] +2\left[ \mathcal{K}^{3}\right] ,  \notag \\
\mathcal{U}_{4} &=&\left[ \mathcal{K}\right] ^{4}-6\left[ \mathcal{K}^{2}%
\right] \left[ \mathcal{K}\right] ^{2}+8\left[ \mathcal{K}^{3}\right] \left[
\mathcal{K}\right] +3\left[ \mathcal{K}^{2}\right] ^{2}-6\left[ \mathcal{K}^{4}\right] ,
\end{eqnarray}%
where $\mathcal{K}_{\nu }^{\mu }=\sqrt{g^{\mu \sigma}f_{\sigma \nu}}$, $f$ is the non-dynamical reference metric (``fiducial metric'') and the rectangular bracket denotes the traces, namely $\left[ \mathcal{K}\right] =%
\mathcal{K}_{\mu }^{\mu }$ and $\left[ \mathcal{K}^{n}\right] =\left(\mathcal{K}^{n}\right) _{\mu }^{\mu }$. 

Varying the action \eqref{action} we have the field equation of this theory as
\begin{eqnarray}
G_{\mu \nu }+\Lambda g_{\mu \nu }+H_{\mu \nu }+m^{2}\chi _{\mu \nu }=0,
\label{Field equation}
\end{eqnarray}
where $G_{\mu \nu }$ is the usual Einstein tensor, and the $H_{\mu \nu} $ and massive tensor $ \chi_{ \mu \nu} $ have following explicit expressions, respectively: 
\begin{eqnarray}
H_{\mu \nu }& =&-\frac{\alpha }{2}\left[ 8R^{\rho \sigma }R_{\mu \rho \nu
\sigma }-4R_{\mu }^{\rho \sigma \lambda }R_{\nu \rho \sigma \lambda
}-4RR_{\mu \nu }+8R_{\mu \lambda }R_{\nu }^{\lambda }\right.  \nonumber  \\
&+&\,\,\left. g_{\mu \nu }\left( R_{\mu \nu \gamma \delta }R^{\mu \nu \gamma
\delta }-4R_{\mu \nu }R^{\mu \nu }+R^{2}\right) \right],\\
\chi _{\mu \nu }& =&-\frac{c_{1}}{2}\left( \mathcal{U}_{1}g_{\mu \nu }-
\mathcal{K}_{\mu \nu }\right) -\frac{c_{2}}{2}\left( \mathcal{U}_{2}g_{\mu
\nu }-2\mathcal{U}_{1}\mathcal{K}_{\mu \nu }+2\mathcal{K}_{\mu \nu
}^{2}\right) \nonumber  \\
&-&\,\,\frac{c_{3}}{2}(\mathcal{U}_{3}g_{\mu \nu }-3\mathcal{U}_{2}
\mathcal{K}_{\mu \nu } +6\mathcal{U}_{1}\mathcal{K}_{\mu \nu }^{2}-6\mathcal{K}_{\mu \nu }^{3})\nonumber \\
&-&\,\,\frac{c_{4}}{2}(\mathcal{U}_{4}g_{\mu \nu }-4\mathcal{U}_{3}\mathcal{K}_{\mu
\nu }+12\mathcal{U}_{2}\mathcal{K}_{\mu \nu }^{2}-24\mathcal{U}_{1}\mathcal{K
}_{\mu \nu }^{3}+24\mathcal{K}_{\mu \nu }^{4}). 
\end{eqnarray}

Using the same choice for the nondynamical reference metric as \cite{Cai},
\begin{equation}
f_{ab}=\text{diag}(0,0,c^{2},c^{2}\sin ^{2}\theta ),  \label{reference}
\end{equation}
where $c$ is a constant with dimension of length. When we take the limit $d\rightarrow 4$, we can obtain the black hole metric as
\begin{equation}
\text{d}s^{2}=-f(r)\text{d}t^{2}+\frac{\text{d}r^{2}}{f(r)}+r^{2}\left(
\text{d}\theta ^{2}+\sin ^{2}\theta \text{d}\varphi ^{2}\right),
\label{metric}
\end{equation}
where
\begin{eqnarray}
f(r)=1+\frac{r^2}{2\alpha}\left(1-\sqrt{1+4\alpha\left(\frac{2M}{r^3}-\frac{1}{\ell^2}-\frac{\gamma}{r}-\frac{\varepsilon}{r^2}\right)}\,\right).
\label{metric}
\end{eqnarray}
We have already set $\gamma=mc c_1/2$ and $\varepsilon=m^2c^2c_2$ to be consistent with the convention in \cite{Hou:2020yni}\footnote{Note that our action here is not exactly the same as in the \cite{Hou:2020yni}. In particular, our cosmological constant are introduced by hand, whereas in \cite{Hou:2020yni} the effective cosmological constant term  emerged from the graviton mass. But this does not matter, because the final form of the metric and the black hole thermodynamics are still the same.}.

Since we are considering the asymptotically AdS case, the black hole event horizon $r_+$ is defined as the largest root of $f(r)=0$. We can write the black hole mass as the function of $r_+$, which yields
\begin{equation}
M=\frac{1}{2r_+}\left(\frac{r_+^4}{\ell^2}+\gamma r_+^3+(\varepsilon+1)r_+^2+\alpha\right),
\label{mass}
\end{equation}
and the Hawking temperature is
\begin{equation}
	T =\left. \frac{f'(r)}{4\pi}\right |_{r=r_+} 
	= \frac{1}{4\pi r_+(r_+^2+2\alpha)}\left(\frac{3r_+^4}{\ell^2}+2\gamma r_+^3+(\varepsilon+1)r_+^2-\alpha \right)
\end{equation}

Once we know the mass and temperature of the black hole and the cross section of the Hawking radiation, we can calculate the evaporation evolution of the black hole using the Stefan-Boltzmann law. Next, we will analyze the cross section and its effect on the evaporation of an infinitely large black hole in AdS spacetime.

\section{Cross section and emission rate of an infinitely large AdS black hole}
\label{section3}

The massive particles cannot reach infinity in the AdS spacetime, thus only the massless particles contribute to the evaporation. By applying geometrical optics approximation, we assume that all the emitted massless particles move along null geodesics \cite{Page:2015rxa}. Once we orient the angular coordinate $\varphi$ and normalize the affine parameter $\lambda$, we have the geodesic equation of the massless particles written as
\begin{align}
\bigg(\frac{\mathrm{d}r}{\mathrm{d}\lambda}\bigg)^2=E^2-J^2\frac{f(r)}{r^2},
\end{align}
where $E=f(r)\frac{\mathrm{d}t}{\mathrm{d}\lambda}$ and $J=r^2\frac{\mathrm{d}\theta}{\mathrm{d}\lambda}$ are the energy and the angular momentum of the emitted particles respectively. If there is a turning point satisfying $\big(\frac{\mathrm{d}r}{\mathrm{d}\lambda}\big)^2=0$, the particle will turn back towards the black hole and thus cannot be detected by an observer on the AdS boundary. Defining $b\equiv {J}/{E}$, the emitted particle can reach infinity only if
\begin{align}
\frac{1}{b^2}> \frac{f(r)}{r^2},
\end{align}
for all $r> r_+$. The impact factor $b_c$ is defined by the maximal value of ${f(r)}/{r^2}$. According to Stefan-Boltzmann law \cite{Vos1989}, the Hawking emission rate is
\begin{align}
\frac{\mathrm{d} M}{\mathrm{d}t}=- Cb_c^{2} T^4,
\label{law}
\end{align}
with the constant $C=\frac{\pi^3 k^4}{15c^3 \hbar^3}$. The $b_c^2$ and the $T^4$ correspond to the cross section and  energy density respectively. We are only concerned about the qualitative features of the evaporation, so without loss of generality, we will set $C$ to be unity: $C=1$.

To understand the exact evolution, we need to solve the $b_c$ and find the appropriate interval of integration. Once we have the form of $f(r)$, we can try to calculate the maximal value of ${f(r)}/{r^2}$ and obtain the $b_c$. At this point we meet two cases. The first one is that the maximal value is located at $r_p$, which is the radius of an unstable circular photon orbit and satisfies the $\left.\frac{\partial}{\partial r}\frac{f(r)}{r^2}\right |_{r=r_p}=0$, thus the impact factor is given by $b_c= {r_p}/{\sqrt{f(r_p)}}$. The second one is $\frac{f(r)}{r^2}$ is a monotonically increasing function of $r$ outside the black hole, or although $r_p$ exists, ${r_p}/{\sqrt{f(r_p)}}$ is smaller than ${r}/{\sqrt{f(r)}}$ at infinity, so the impact factor $b_c$ will be the effective AdS radius. Thus, we must first determine if there is an unstable photon orbit with radius $r_p$ in the corresponding metric, and if so, we must compare the corresponding value of $f(r)/r^2$ at $r_p$ with the value at infinity to obtain the correct form of $b_c$. See the case of massive gravity \cite{Hou:2020yni} for an example. 

{The above analysis is easy to do when the formula of $f(r)$ is simple. However, as it becomes more complicated, which is often accompanied by changes in qualitative properties due to the increase of parameters, our calculations can involve the existence of roots of higher degree equations and specific numerical values. For example, in the case of Gauss-Bonnet massive gravity, $\frac{\partial}{\partial r}\frac{f(r)}{r^2}=0$ will yield a quintic equation, which is obviously much harder than the quadratic equation in massive gravity. We can naturally deduce that this could be worse in other, more complicated gravity models.}

Now the problem arises as to which variables determine the qualitative features of the lifetime of an infinitely large black hole. We already know that the lifetime of a black hole in four-dimensional asymptotically flat spacetime in Einstein gravity is $t\sim M_0^3$, so it will diverge if we take an infinite initial mass. Although the temperature of the black hole rises as it loses mass, it still cannot evaporate an infinite amount of mass in a finite amount of time. On the other hand, for the AdS black hole in massive gravity, the black hole can lose infinite mass in finite time, but the black hole may have a zero temperature state, and thus, according to the third law of thermodynamics, the black hole will never evaporate out and become a remnant in late time. So, in order to obtain the qualitative features of the life of an infinitely large AdS black hole, we need to answer the following two questions:

1. Could an infinitely large AdS black hole evaporate an infinite amount of mass in a finite amount of time?

2. Are there zero temperature states that ultimately prevent the evaporation?

In this section we will answer the first question by proving that \emph{an infinitely large AdS black hole is indeed capable of evaporating an infinite amount of mass in finite time}.

For an infinitely large four-dimensional AdS black hole, its metric can be roughly written as
\begin{eqnarray}
f(r)=1+\frac{r^2}{\ell_{\text{eff}}^2}+\mathcal{O}(r,\text{const},r^{-1},...).
\end{eqnarray}
Since the metric is asymptotically AdS, we have the effective AdS radius $\ell_{\text{eff}}^2>0$, although $\ell_{\text{eff}}$ may be corrected so that it does not correspond to the cosmological constant in the action. In our case we have $\ell_{\text{eff}}^2=\frac{\ell^2}{2}\left(1 +\sqrt{1-\frac{4\alpha}{\ell^2}}\right)$. 

On the other hand, the black hole mass \eqref{mass} satisfies $M\sim r_+^3$ as $r_+\rightarrow \infty$, so the terms related to $M$ in the metric can be written in a form of $M^n/r^{3n-2}$, where $n$ is an integer. In addition, we also have a linear term of $r$ corresponding to the $\gamma$ in the metric \eqref{metric}. 

Next we will prove that \emph{the impact factor of an infinitely large AdS black hole is $b_c=\ell_{\text{eff}}$.} Some readers may think this is obvious, because $r_p/\sqrt{1+\frac{r_p^2}{\ell_{\text{eff}}^2}}$ equals $\ell_{\text{eff}}$ in $r_p\rightarrow \infty$. However, we need to note that $M$ in the metric is also infinite at this point, and corresponding terms also need to be compared with $r_p^2$ of order. To better illustrate the issue, we first consider a simplified $f(r)$ such that
\begin{eqnarray}
f(r)=1+\frac{r^2}{\ell_{\text{eff}}^2}+ar-\frac{M^n}{r^{3n-2}},
\end{eqnarray}
where $a$ is a positive coefficient corresponding to the contribution of $\gamma$. Solving the equation $\frac{\partial}{\partial r}\frac{f(r)}{r^2}=0$ we have \footnote{Here we assume that $r_p$ must exists. If $r_p$ does not exist, the impact factor is directly $\ell_{\text{eff}}$.}
\begin{eqnarray}
ar^{3n-1}+2r^{3n-2}-3nM^n=0
\end{eqnarray}
For $M\rightarrow \infty$ we have the solution $r_p\sim M^{\frac{n}{3n-1}}\sim r_+^{\frac{3n}{3n-1}}$, and the $r_p \sim r_+ ^{3/2}$ as $n=1$ and $r_p \sim r_+$ as $n\rightarrow \infty$ \footnote{In the Schwarzschild AdS case, we have $a=0$ and $n=1$, so the $r_p\sim M \sim r_+^3$.}. We can add the corresponding coefficients and more related $M$ terms, and all these contributions of $M$ terms will make $r_p \sim r_+^m$, where $1<m<\frac{3}{2}$. Thus, all the $M$ related terms in $f(r_p)$ can be written as 
\begin{eqnarray}
\frac{M^n}{r_p^{3n-2}}\sim \frac{r_+^{3n}}{r_+^{m(3n-2)}}\sim r_+^{2m-(m-1)3n},
\end{eqnarray}

Since $m>1$, the above term is always smaller than the $r_+^{2m}$ of the ${r_p^2}/{\ell_{\text{eff}}^2}$ term. Thus we can conclude that the leading order term in $f(r_p)$ is the ${r_p^2}/{\ell_{\text{eff}}^2}$ term and we have $b_c= {r_p}/{\sqrt{f(r_p)}}=\ell_{\text{eff}}$.

Defining the dimensionless variables $x\equiv {r_+}/{\ell}$, we have
\begin{eqnarray}
M\sim x^3 \ell,\quad  T\sim x \ell^{-1},
\end{eqnarray}
and from the Stefan-Boltzmann law we know
\begin{eqnarray}
\mathrm{d}t \sim -\frac{\mathrm{d} M}{b_c^2 T^4} \sim -\frac{x^2 \ell}{\ell_{\text{eff}}^2 x^4 \ell^{-4}} \mathrm{d}x\sim -\frac{\ell^5}{\ell_{\text{eff}}^2}\frac{1}{x^2}\mathrm{d} x.
\end{eqnarray}
The integral of the above equation is naturally convergent at infinity, so we can obtain an infinitely large AdS black hole capable of evaporating an infinite amount of mass in finite time. In the Schwarzschild AdS spacetime we have $\ell_{\text{eff}}=\ell$, and the result reduces to the Page's conclusion that the total Hawking decay time is bounded by a time of the order of $\ell^3$ \cite{Page:2015rxa}.

All the above analyses can be extended to the general dimension $d\geq 4$. For general $d$ we have $T\sim r_+$ and $M\sim r_+^{d-1}$. We can also prove that the impact factor remains $\ell_{\text{eff}}$ and the time integral in the Stefan-Boltzmann law converges at infinity.

\section{The final state: $T=0$ or $T=\infty$?}
\label{section4}

In the previous section we have proved that an infinitely large AdS black hole can evaporate an infinite amount of mass in finite time. However, this does not mean that it can evaporate all the time, because we also have to consider the temperature behavior and the emission rate of the black hole as it gets smaller. This brings us back to the second question: are there zero temperature states that ultimately prevent evaporation? If so, the final state of the black hole will be a remnant, otherwise it will have completely evaporated.

There are only two cases of final state temperature in black hole evaporation. The first one is $T=0$, which also corresponds to $\frac{\partial M}{\partial r_+}=0$. The other one is $T=\infty$, which corresponds to $r_+=\sqrt{-2\alpha}$. This case is only possible in $\alpha<0$. If both cases occur, we must compare the radius of the corresponding black holes and check the evaporation path to determine which is the true final state.

In principle, we could just solve the equation $T = 0$ and find all possible cases, just like in the case of massive gravity \cite{Hou:2020yni}. {However, we still encounter problems with higher degree equations.} In massive gravity, $T=0$ corresponds to a quadratic equation and we can solve it analytically, but in our case it is a quartic equation, so the analysis of the existence of roots and other related properties would be very complicated. Therefore, instead of solving the quartic equation directly, we will start from the known properties of massive gravity in \cite{Hou:2020yni} and treat the Gauss-Bonnet term as a correction\footnote{{We could also get the Gauss-Bonnet result first and then introduce massive gravity as a correction. However, the massive gravity itself is more complicated and has several parameters, while there is only one parameter $\alpha$ in the Gauss-Bonnet gravity, so it is more convenient to use the Gauss-Bonnet term as a correction.}}. {We expect to obtain direct qualitative information about the temperature by making corrections based on known result.} With different choices of parameters, the Gauss-Bonnet term affects the thermodynamics of black holes in different ways. We can classify the final state temperature in the evaporation process and determine whether the black hole has a finite lifetime.

\subsection{$\gamma^2\ell^2<3(\varepsilon+1)$, or $\gamma^2\ell^2>3(\varepsilon+1)$, $\gamma>0$, $\varepsilon+1>0$}

For $\alpha=0$, these two cases correspond to $T=0$ having no real positive root\footnote{For $\alpha=0$, in $\gamma^2\ell^2<3(\varepsilon+1)$, the roots are not real, and in $\gamma^2\ell^2>3(\varepsilon+1)$, $\gamma>0$, $\varepsilon+1>0$, the roots are negative \cite{Hou:2020yni}.}, meaning in both cases $M$ is a monotonically increasing function of $r_+$ and thus there is no $T=0$ state. However, $\alpha \neq 0$ will change the asymptotic behavior of $M$. The $M\rightarrow 0$ as $r_+\rightarrow 0$ in $\alpha =0$, but the $M\rightarrow +\infty$ as $r_+\rightarrow 0$ in $\alpha >0$, and the $M\rightarrow -\infty$ as $r_+\rightarrow 0$ in $\alpha <0$ \footnote{The Gauss-Bonnet coefficient $\alpha$ is supposed to be postive in string theory\cite{Boulware:1985wk}, but for generality we consider both cases of $\alpha>0$ and $\alpha<0$ in the present work. See e.g. \cite{Ong:2022mmm} for more discussion in the issue.}. 

Note that this does not mean that $M$ can be arbitrarily small in $\alpha <0$. The $M$ is solved from the equation $f(r_+)=0$, and there is a square root in $f(r)$, so if $M$ becomes small enough it can make the square root term negative, thus making the metric $f(r)$ unphysical. We can see that when the square root term is zero, there is $r_+ = \sqrt{-2\alpha}$, which corresponds to $T = \infty$. This is the smallest radius the black hole can take. If we continue to reduce the corresponding $M$, the metric will not be well defined.

\begin{figure}[h!]
	\begin{center}
		\includegraphics[width=0.40\textwidth]{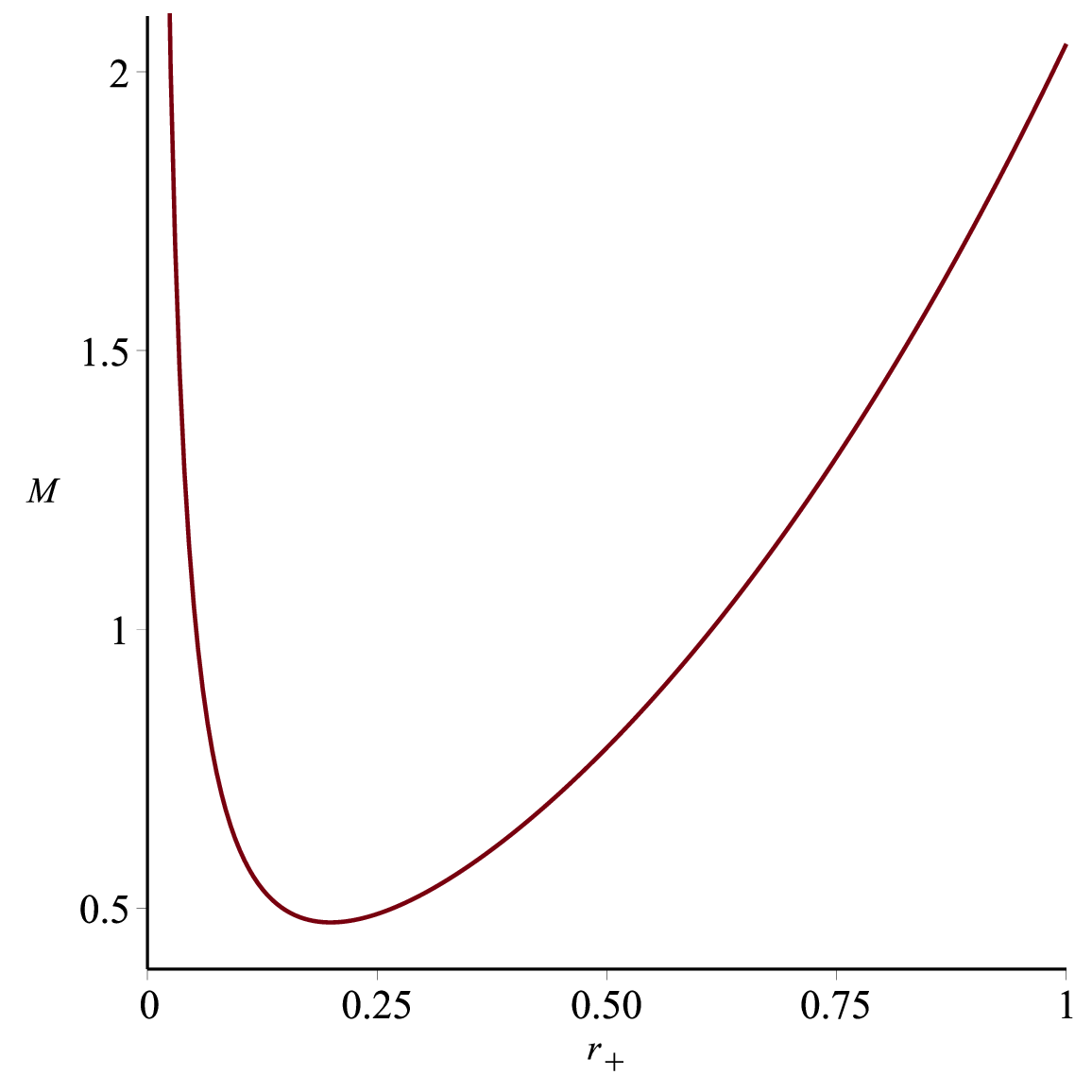}
		\includegraphics[width=0.40\textwidth]{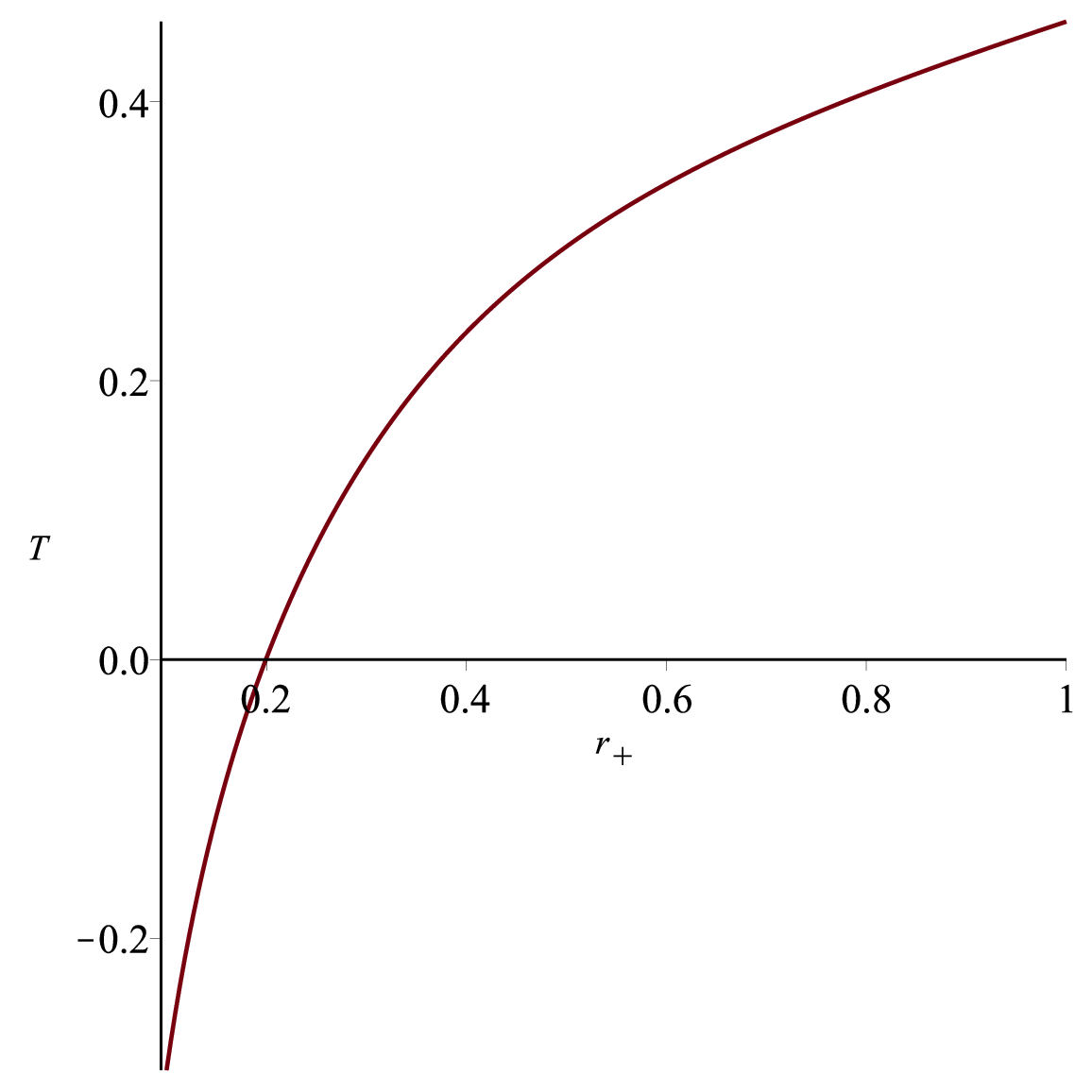}
		\vspace{-1mm}
		\caption{The black hole mass $M$ and temperature $T$ as function of the radius $r_+$ with setting $\gamma=1$, $\ell=1$, $\varepsilon=1$ and $\alpha=0.1$.  }
		\label{fig1}
	\end{center}
\end{figure}

In Figure.\ref{fig1} we present an example of the black hole mass $M$ and temperature $T$ as function of the radius $r_+$ with setting $\gamma=1$, $\ell=1$, $\varepsilon=1$ and $\alpha=0.1$. Since $M$ is divergent at both $r_+$ tends to zero and infinity, it must have a minimal value corresponding to $T = 0$. This also means that the black hole finishes evaporating and the final state of the black hole is a remnant. Note that in all the graphs in this paper, we present the whole curve as the function of $r_+$ for clarity, but \emph{negative part of $T$ (and the corresponding part of $M$) is not physical and should be excluded. If the values of $M$ correspond to multiple $r_+$, only the largest $r_+$ is the real black hole horizon, and the others are the inner horizons. }

\begin{figure}[h!]
	\begin{center}
		\includegraphics[width=0.40\textwidth]{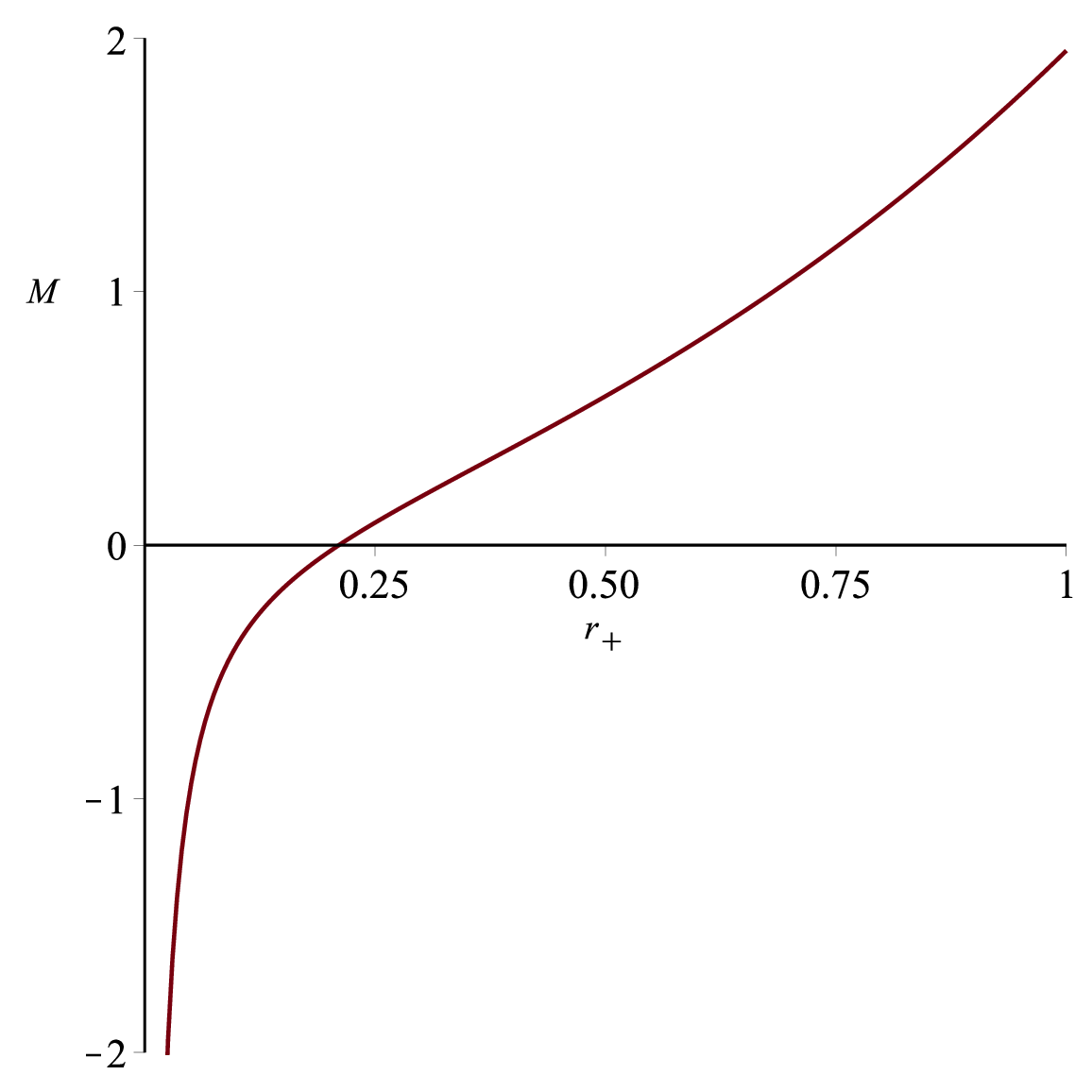}
		\includegraphics[width=0.40\textwidth]{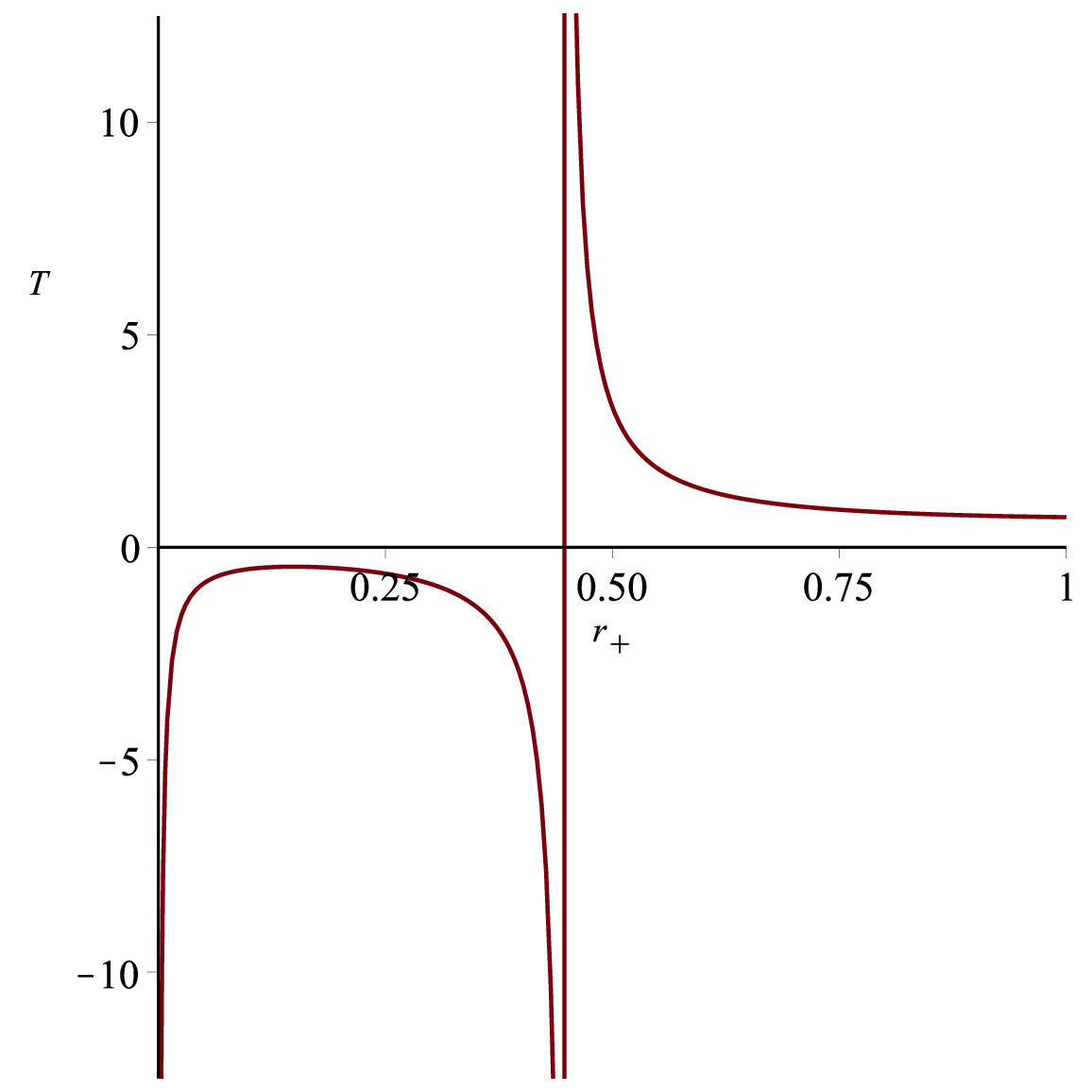}
		\vspace{-1mm}
		\caption{The black hole mass $M$ and temperature $T$ as function of the radius $r_+$ with setting $\gamma=1$, $\ell=1$, $\varepsilon=1$ and $\alpha=-0.1$.  }
		\label{fig2}
	\end{center}
\end{figure}

In Figure.\ref{fig2} we present an example of the black hole mass $M$ and temperature $T$ as function of the radius $r_+$ with setting $\gamma=1$, $\ell=1$, $\varepsilon=1$ and $\alpha=-0.1$. The $M$ is still a monotonically increasing function of $r_+$, but the $M\rightarrow -\infty$ as $r_+\rightarrow 0$. From the graph of $T(r_+)$ we know the the temperature is divergent at $r_+=\sqrt{-2\alpha}$ , so we know that the black hole can evaporate out in finite time.

\subsection{$\gamma^2\ell^2>3(\varepsilon+1)$, $\varepsilon+1<0$}

\begin{figure}[h!]
	\begin{center}
		\includegraphics[width=0.40\textwidth]{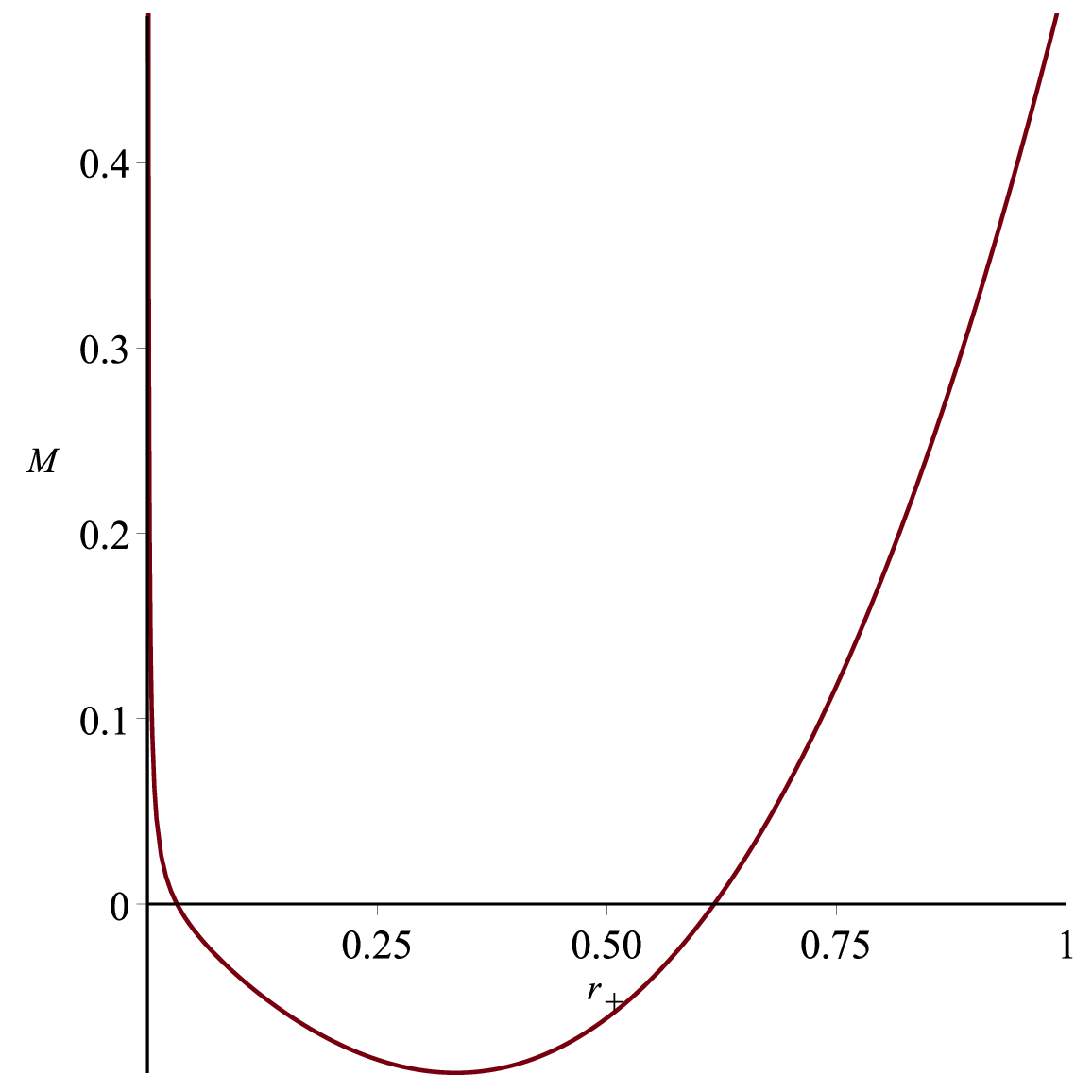}
		\includegraphics[width=0.40\textwidth]{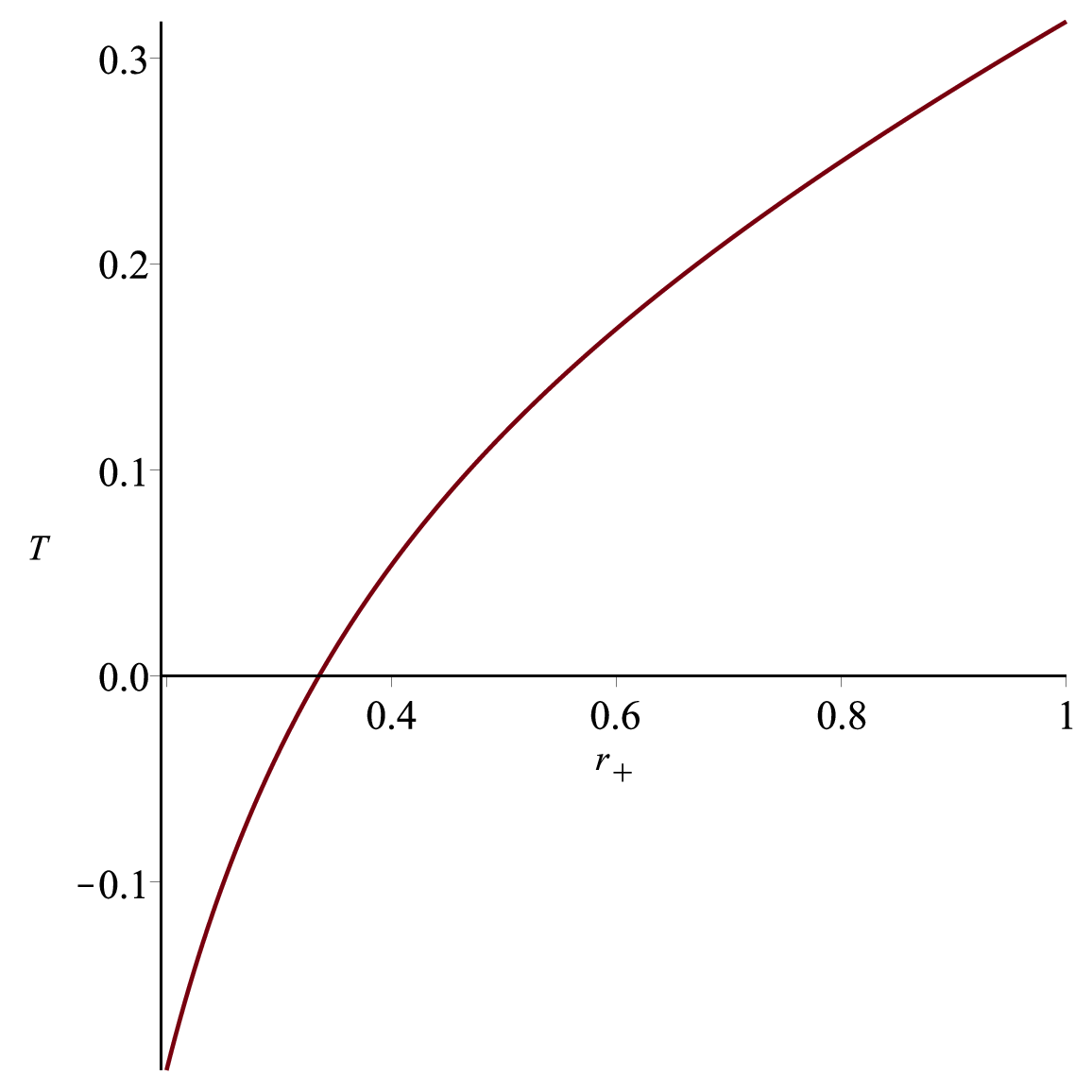}
		\vspace{-1mm}
		\caption{The black hole mass $M$ and temperature $T$ as function of the radius $r_+$ with setting $\gamma=1$, $\ell=1$, $\varepsilon=-2$ and $\alpha=0.001$.  }
		\label{fig3}
	\end{center}
\end{figure}

In this case when $\alpha=0$ the $M$ has already a minimal value, and as $\alpha>0$ only the asymptotic behavior of $M$ at $r_+\rightarrow 0$ will change, so we will still have a minimal value corresponding to $\frac{\partial M}{\partial r_+}=0$. The final state of the black hole is still $T = 0$. In Figure.\ref{fig3} we present an example of black hole mass $M$ and temperature $T$ as function of the radius $r_+$ with setting $\gamma=1$, $\ell=1$, $\varepsilon=-2$ and $\alpha=0.001$. 

\begin{figure}[h!]
	\begin{center}
		\includegraphics[width=0.50\textwidth]{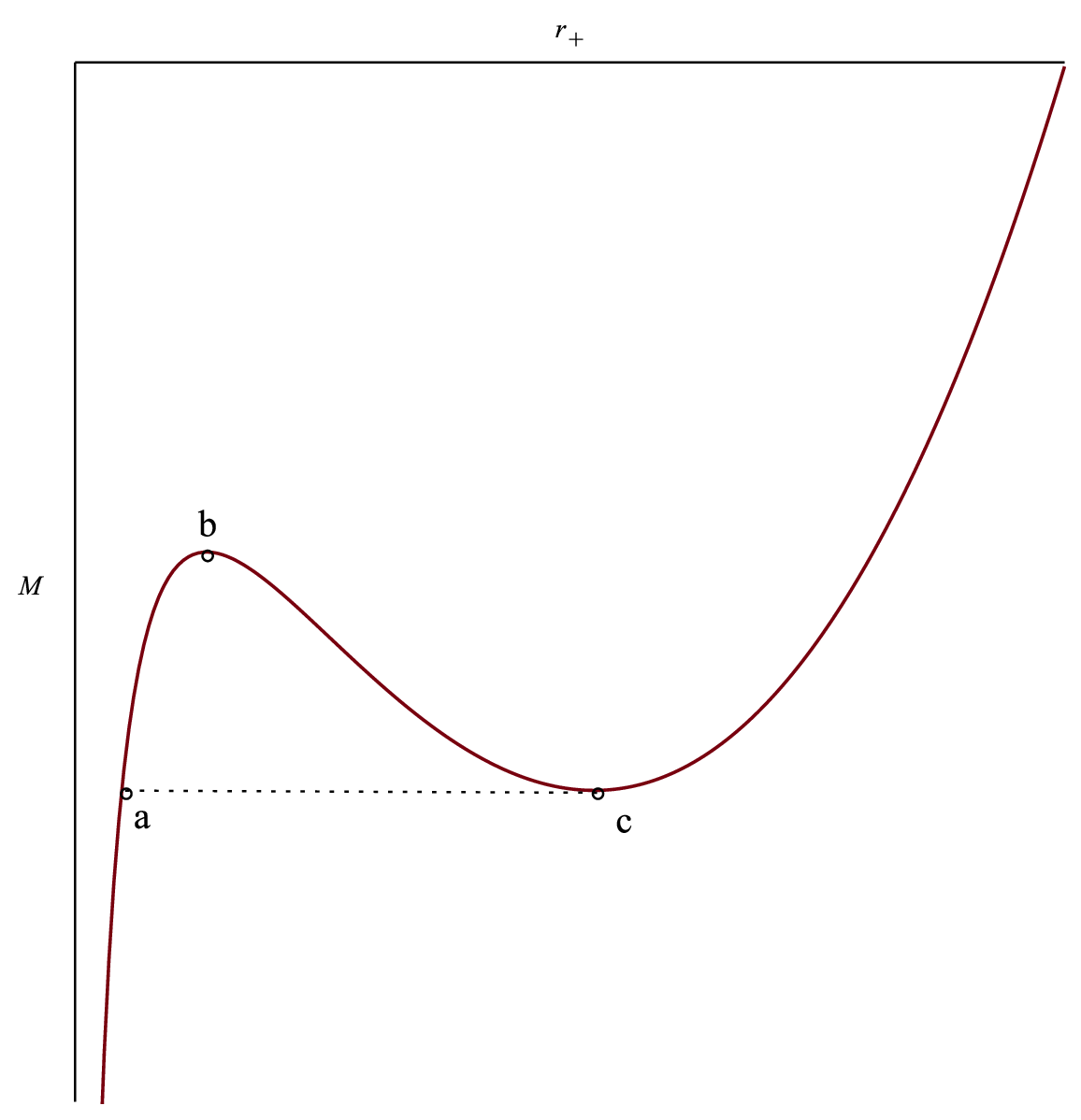}
		\vspace{-1mm}
		\caption{Schematic diagram of $M$ as $\alpha<0$ in $\gamma^2\ell^2>3(\varepsilon+1)$, $\varepsilon+1<0$.}
		\label{fig4}
	\end{center}
\end{figure}

On the other hand, when $\alpha<0$, the curve of $M$ is as shown in the schematic we present in Figure.\ref{fig4}. We will have two extreme points b and c in the figure, while point a corresponds to the same $M$ as point c. Now the question is where is the position of $\sqrt{-2\alpha}$?

Roughly speaking we have four possible situations: $\sqrt{-2\alpha}>r_c$; $r_b<\sqrt{-2\alpha}<r_c$; $r_a<\sqrt{-2\alpha}<r_b$, and $\sqrt{-2\alpha}<r_a$. Of course the two curves ab and bc do not correspond to the event horizon, but to the inner horizon, so even we have $r_b<\sqrt{-2\alpha}<r_c$ or $r_a<\sqrt{-2\alpha}<r_b$, they will not affect the evaporation of the black hole. Nevertheless, we can directly calculate 
\begin{equation}
\left. \frac{\partial M}{\partial r_+}\right |_{r_+=\sqrt{-2\alpha}} 
	= \frac{1}{4\ell^2}\left( 4\sqrt{-2\alpha}\gamma \ell^2 +2\varepsilon\ell^2+3\ell^2-12\alpha  \right).
\label{partial}
\end{equation}
From the above equation, it follows that if we fix $\gamma$, $\varepsilon$, and $\ell$, as long as $|\alpha|$ is large enough (this also means that $\sqrt{-2\alpha}$ is large enough), we will always get the above formula to be positive. If point b and point c still exist at this point \footnote{Under certain parameter choices, points b and c may no longer exist, and $M$ will become a monotonically increasing function of $r_+$. This is similar to the case of $\sqrt{-2\alpha}>r_c$, where the final states of the black hole are $T = \infty$ and the black hole will have completely evaporated.}, we will have $\sqrt{-2\alpha}>r_c$, and the final state of the black hole is $T = \infty$ and the black hole will have completely evaporated.

When we decrease $|\alpha|$, the $\sqrt{-2\alpha}$ can be smaller than $r_c$. Let us see what happens when the $|\alpha|$ is very small. We know that when $\alpha=0$, the branch from point b to $-\infty$ ceases to exist, so the location of point b and a should be the functions of $\alpha$. From eq.\eqref{partial} we can conclude that when $|\alpha|$ is very small, as $2\varepsilon+3<0$, eq.\eqref{partial} will be negative, meaning the $r_b<\sqrt{-2\alpha}<r_c$, and as $2\varepsilon+3>0$, the eq.\eqref{partial} will be positive, meaning the $\sqrt{-2\alpha}<r_b$. 

Our final question is: assuming $\sqrt{-2\alpha}<r_b$, does it satisfy $r_a<\sqrt{-2\alpha}<r_b$, or can we even have $\sqrt{-2\alpha}<r_a$? If there is $\sqrt{-2\alpha}<r_a$, this means that the black hole has a new branch of solutions: from point a to $r_+=\sqrt{-2\alpha}$. As mentioned before, we already know that the position of point a should also be a function of $\alpha$, so we have 
\begin{equation}
\frac{1}{2r_a}\left(\frac{r_a^4}{\ell^2}+\gamma r_a^3+(\varepsilon+1)r_a^2+\alpha\right)=M(r_c).
\label{15}
\end{equation}

When $|\alpha|$ is small enough, $M(r_c)$ approachs to the value in the case of $\alpha=0$, and $r_a$ should also be an infinitesimal with respect to $\alpha$. From the above equation we can deduce that $r_a \sim |\alpha|$. On the other hand, $r_+=\sqrt{-2\alpha}\sim |\alpha|^{1/2}$, which means that when $|\alpha|$ approaches zero, the $r_a$ has to converge to zero faster than $\sqrt{-2\alpha}$. This also implies that we can only have $r_a<\sqrt{-2\alpha}<r_b$, not $\sqrt{-2\alpha}<r_a$ in small $|\alpha|$.

\begin{figure}[h!]
	\begin{center}
		\includegraphics[width=0.45\textwidth]{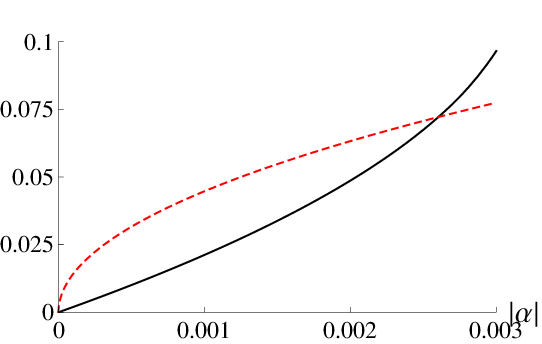}
		\vspace{-1mm}
		\caption{{The positional relationship between $\sqrt{-2\alpha}$ and $r_a$ in $\gamma=1$, $\ell=1$, $\varepsilon=-1.5$. The dashed and solid curves correspond to $\sqrt{-2\alpha}$ and $r_a$ respectively.}}
		\label{fig}
	\end{center}
\end{figure}

{However, the above analysis assumes that $|\alpha|$ is small enough. As we gradually increase the value of $|\alpha|$, we will find that the situation can change instead. In Figure.\ref{fig} we give a numerical result comparing the positional relationship between $\sqrt{-2\alpha}$ and $r_a$ as other parameters are set. The dashed and solid curves correspond to $\sqrt{-2\alpha}$ and $r_a$ respectively. We can see that $\sqrt{-2\alpha}$ is larger than $r_a$ when $|\alpha|$ is very small, and as $|\alpha|$ continues to increase, we can observe intervals where $\sqrt{-2\alpha}$ is smaller than $r_a$. This also means that we have a new interval of evaporation, i.e. from $r_a$ to $\sqrt{-2\alpha}$. As we continue to increase $|\alpha|$, the properties of the $M$ change to monotonically increasing, and we are back to the case of 4.1.}

\begin{figure}[h!]
	\begin{center}
		\includegraphics[width=0.243\textwidth]{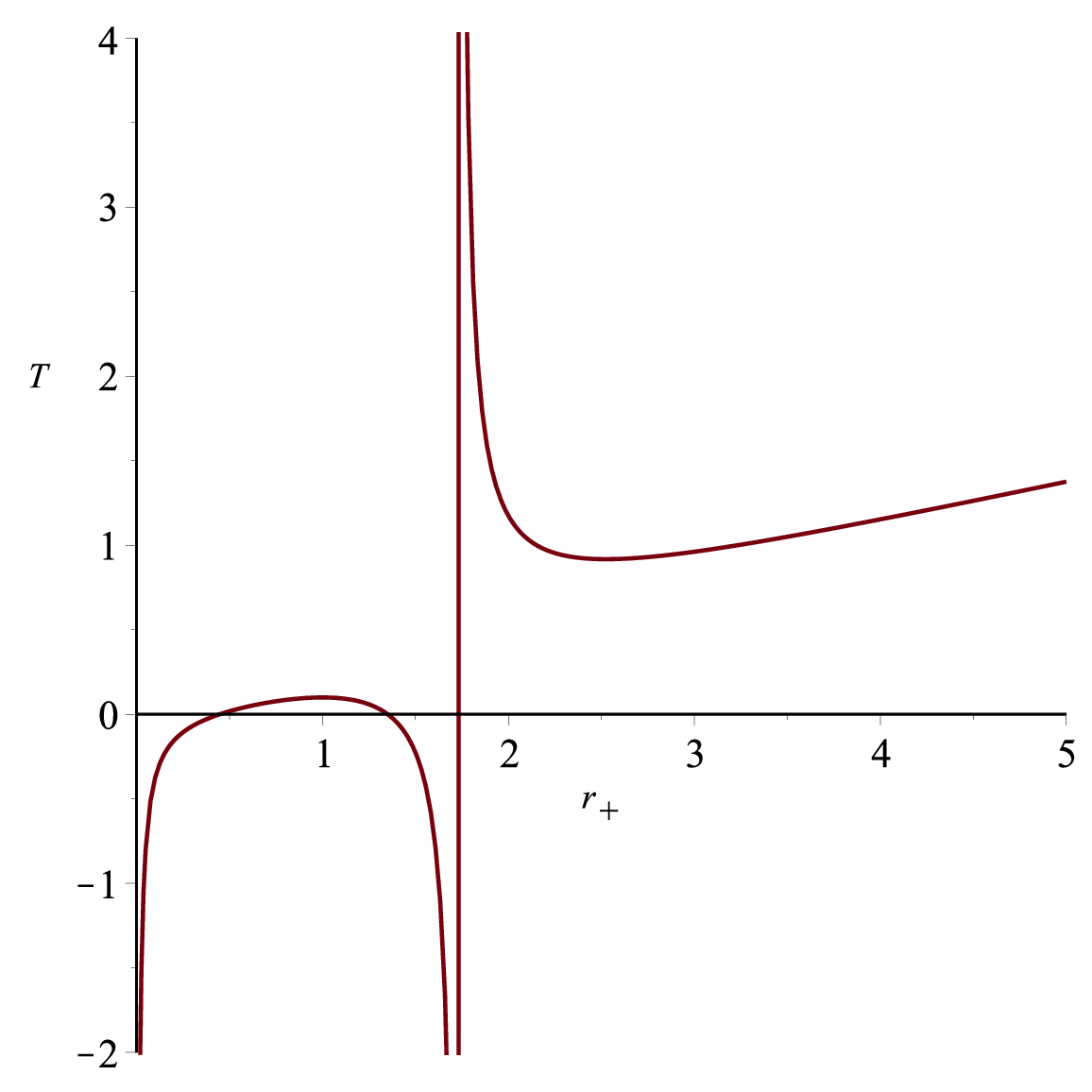}
		\includegraphics[width=0.243\textwidth]{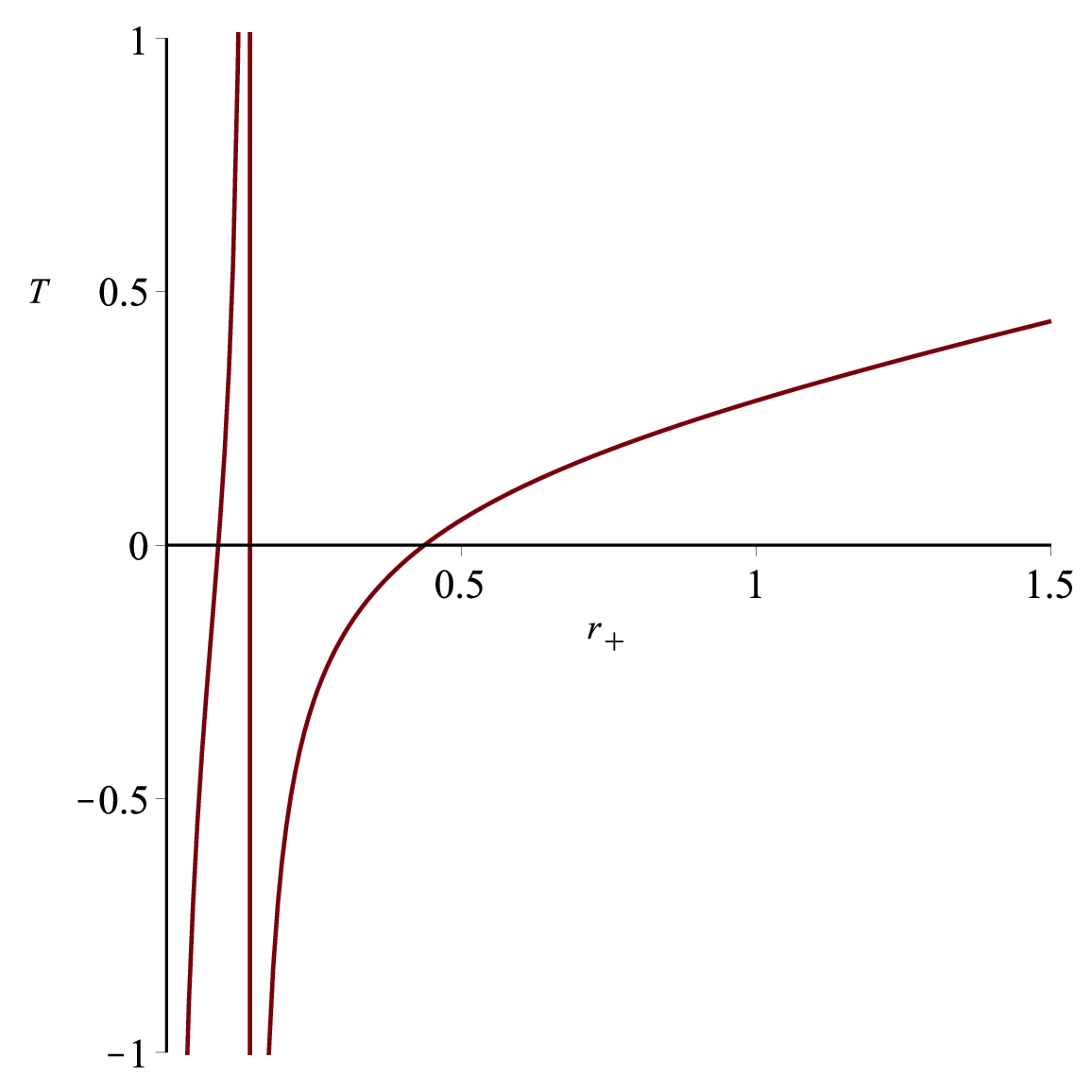}
		\includegraphics[width=0.243\textwidth]{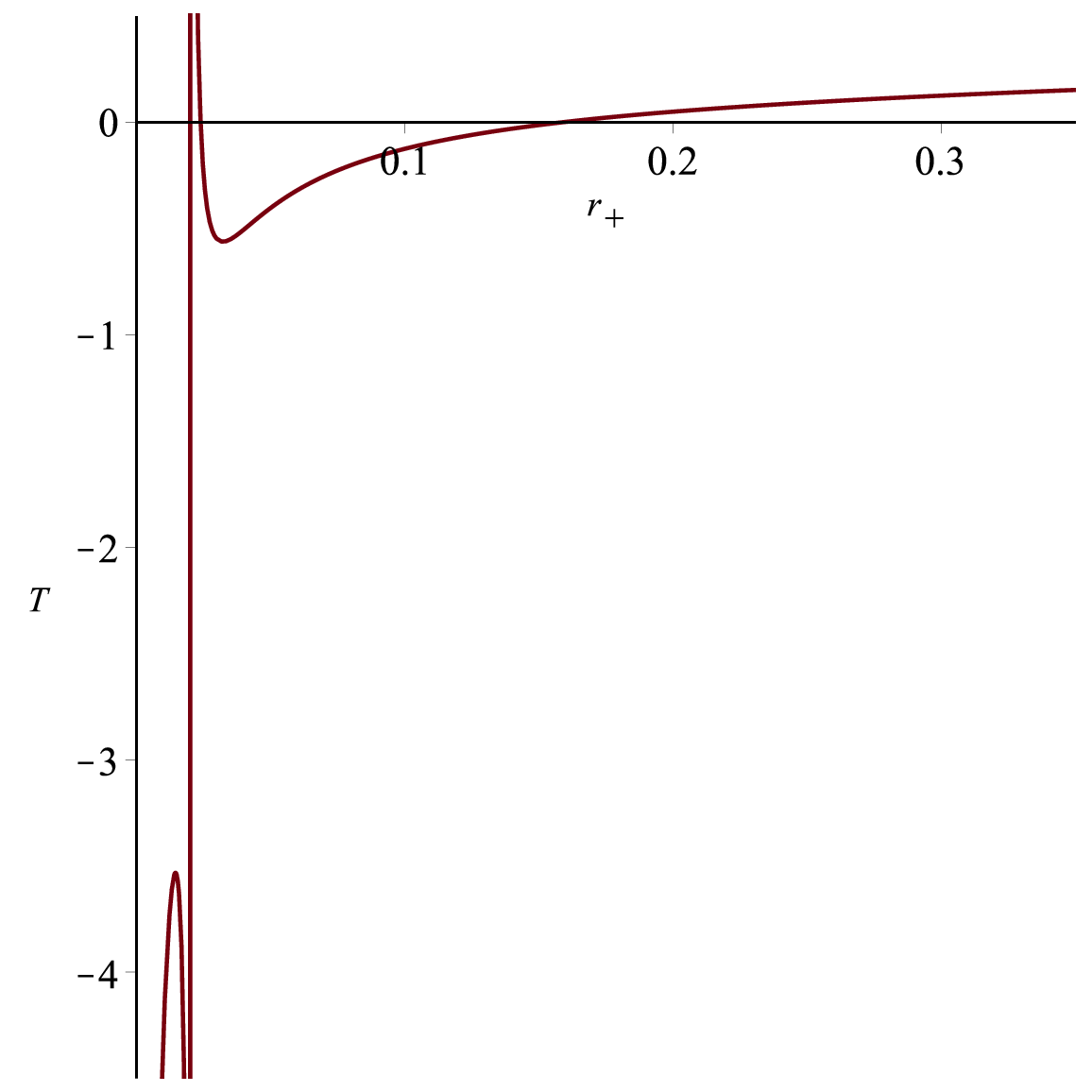}
		\includegraphics[width=0.243\textwidth]{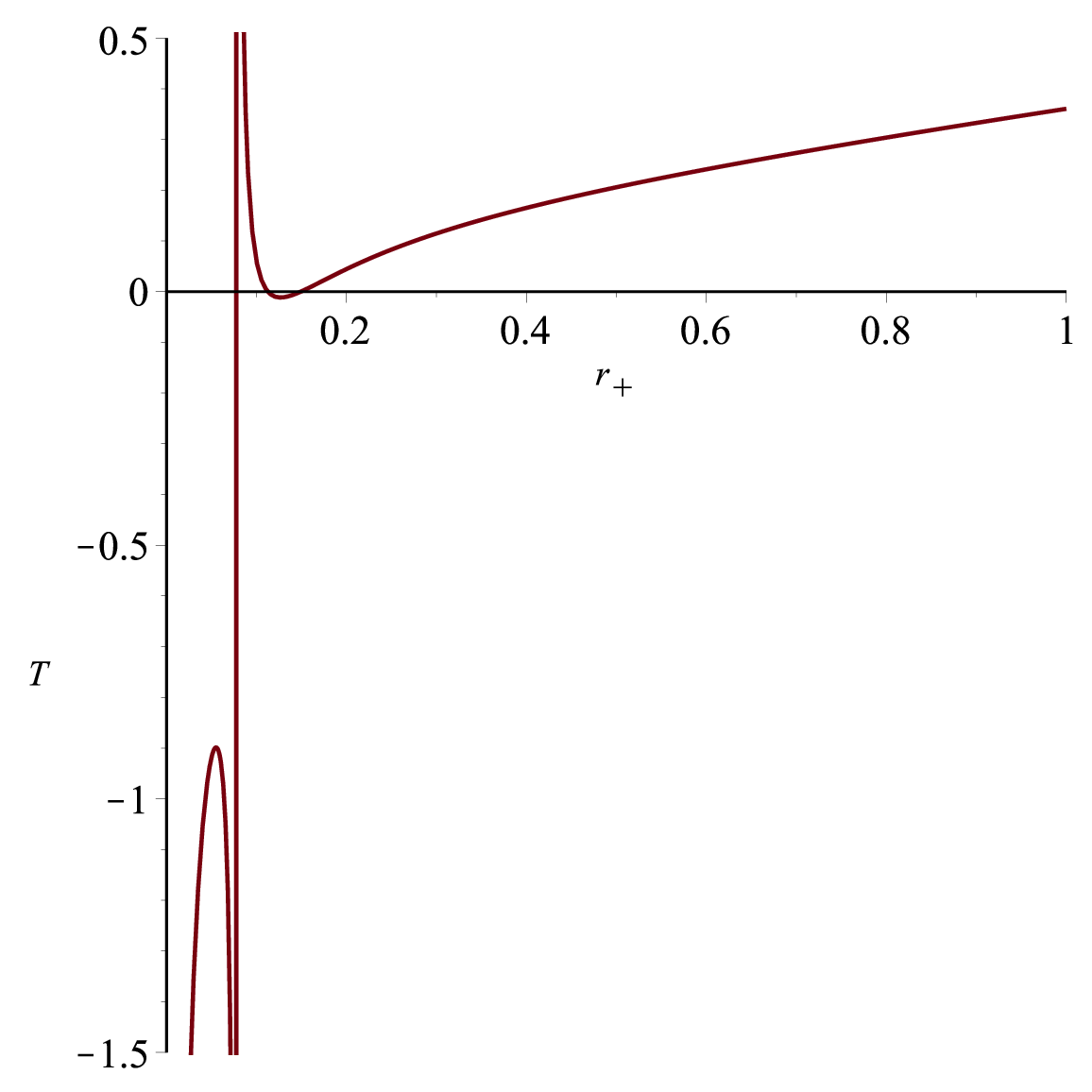}
		\vspace{-1mm}
		\caption{{The black hole temperature $T$ as function of the radius $r_+$. In the first figure we set $\gamma=1$, $\ell=1$, $\varepsilon=-10$, $\alpha=-1.5$. In the second figure we set $\gamma=1$, $\ell=1$, $\varepsilon=-2.5$, $\alpha=-0.01$. In the third figure we set $\gamma=1$, $\ell=1$, $\varepsilon=-1.4$, $\alpha=-0.0002$. In the fourth figure we set $\gamma=1$, $\ell=1$, $\varepsilon=-1.5$, $\alpha=-0.003$. They  correspond to $\sqrt{-2\alpha}>r_c$, $r_b<\sqrt{-2\alpha}<r_c$, $r_a<\sqrt{-2\alpha}<r_b$, and $\sqrt{-2\alpha}<r_a$ respectively.  }}
		\label{fig5}
	\end{center}
\end{figure}

{In Figure.\ref{fig5} we present some examples of the black hole temperature $T$ as the function of $r_+$. From left to right the cases correspond to $\sqrt{-2\alpha}>r_c$, $r_b<\sqrt{-2\alpha}<r_c$, $r_a<\sqrt{-2\alpha}<r_b$, and $\sqrt{-2\alpha}<r_a$ respectively.}

We summarize this subsection by stating that when $\alpha>0$, there is always a $T=0$ state and the black hole will not evaporate completely. When $\alpha<0$, if $|\alpha|$ is large enough to satisfy $\sqrt{-2\alpha}>r_c$, the final state of the black hole can be $T=\infty$ and thus evaporate completely, and if $r_a<\sqrt{-2\alpha}<r_c$, the final state of the black hole is $T=0$ and will not evaporate completely. {Finally, the presence of $\sqrt{-2\alpha}<r_a$ implies a new evaporation path, i.e., point a to $\sqrt{-2\alpha}$.}

\subsection{$\gamma^2\ell^2>3(\varepsilon+1)$, $\varepsilon+1>0$, $\gamma<0$}

\begin{figure}[h!]
	\begin{center}
		\includegraphics[width=0.40\textwidth]{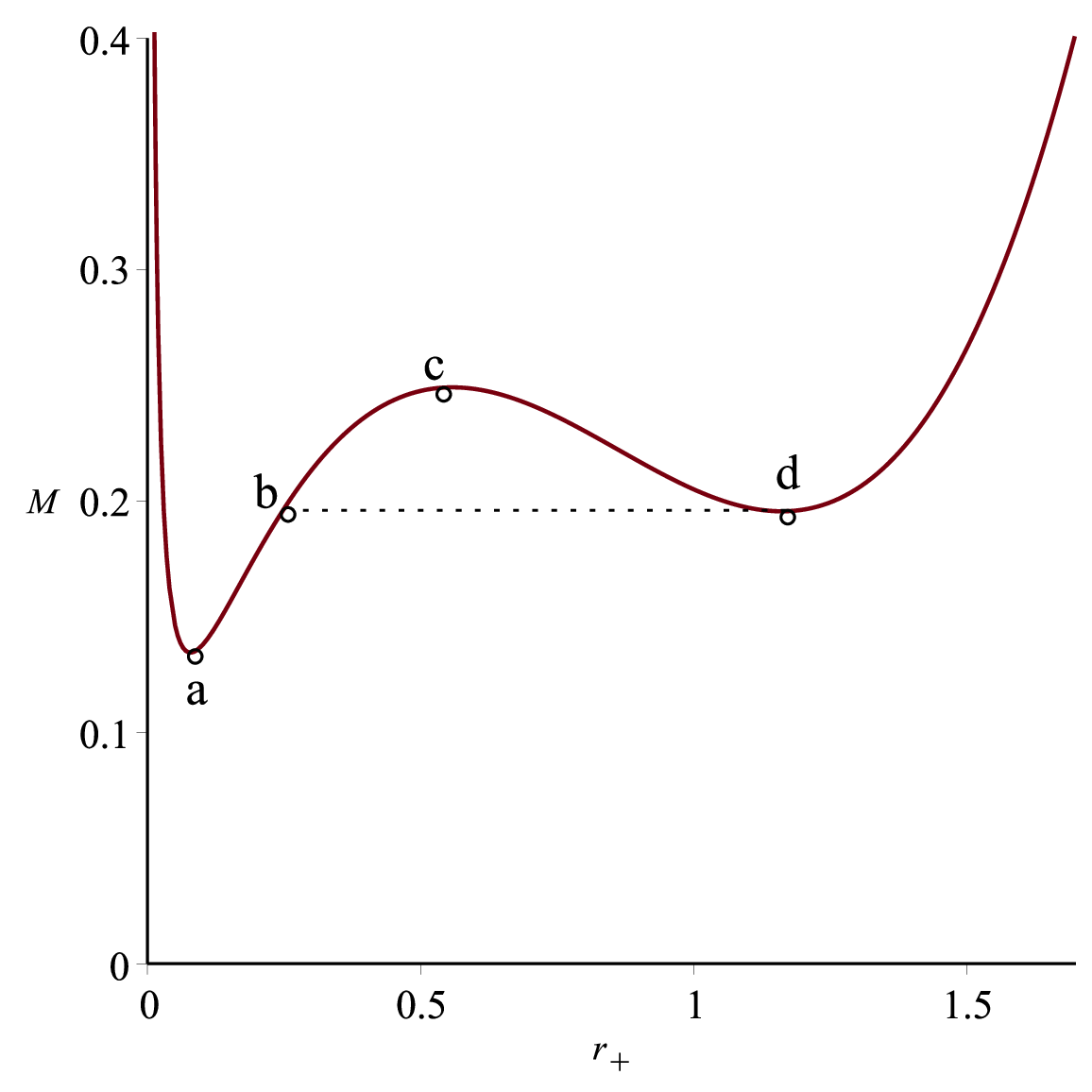}
		\includegraphics[width=0.40\textwidth]{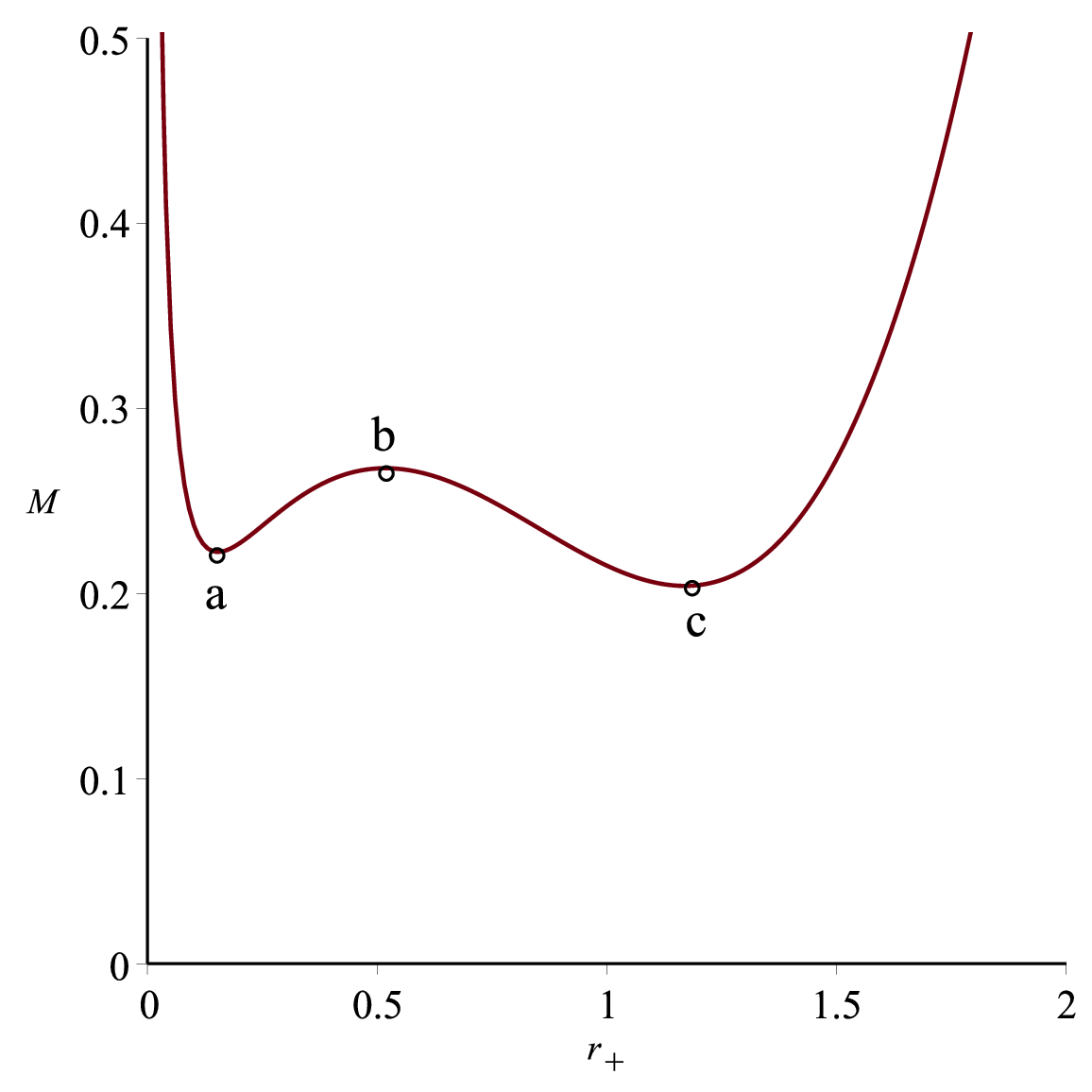}
		\vspace{-1mm}
		\caption{The black hole mass $M$ as function of the radius $r_+$. In the left figure we set $\gamma=1$, $\ell=1$, $\varepsilon=-2.6$, $\alpha=0.01$. In the right figure we set $\gamma=1$, $\ell=1$, $\varepsilon=-2.6$, $\alpha=0.03$.  }
		\label{fig6}
	\end{center}
\end{figure}

In this case, as $\alpha=0$ the $M$ has two minimal values, and as $\alpha>0$ the asymptotic behavior of $M$ at $r_+\rightarrow 0$ will change too. In Figure.\ref{fig6} we present two examples of black hole mass $M$ as function of the radius $r_+$. In the first case we have $M(r_a)<M(r_d)$, which means there are two branches of the black hole solution: from $\infty$ to $r_d$, and from $r_b$ to $r_a$. Both final states ($r_d$ and $r_a$) correspond to $T=0$, so the black hole could not evaporate completely. In the second case we have $M(r_a)>M(r_c)$, which means there is only one branch: from $\infty$ to $r_c$, and the black hole also could not evaporate completely. If we continue to increase the value of $\alpha$, we may only have one minimum value of $M$, and the curve of $M$ is similar to the case in Fig.\ref{fig1}. This does not change the qualitative features of the black hole evaporation and is omitted here.

\begin{figure}[h!]
	\begin{center}
		\includegraphics[width=0.50\textwidth]{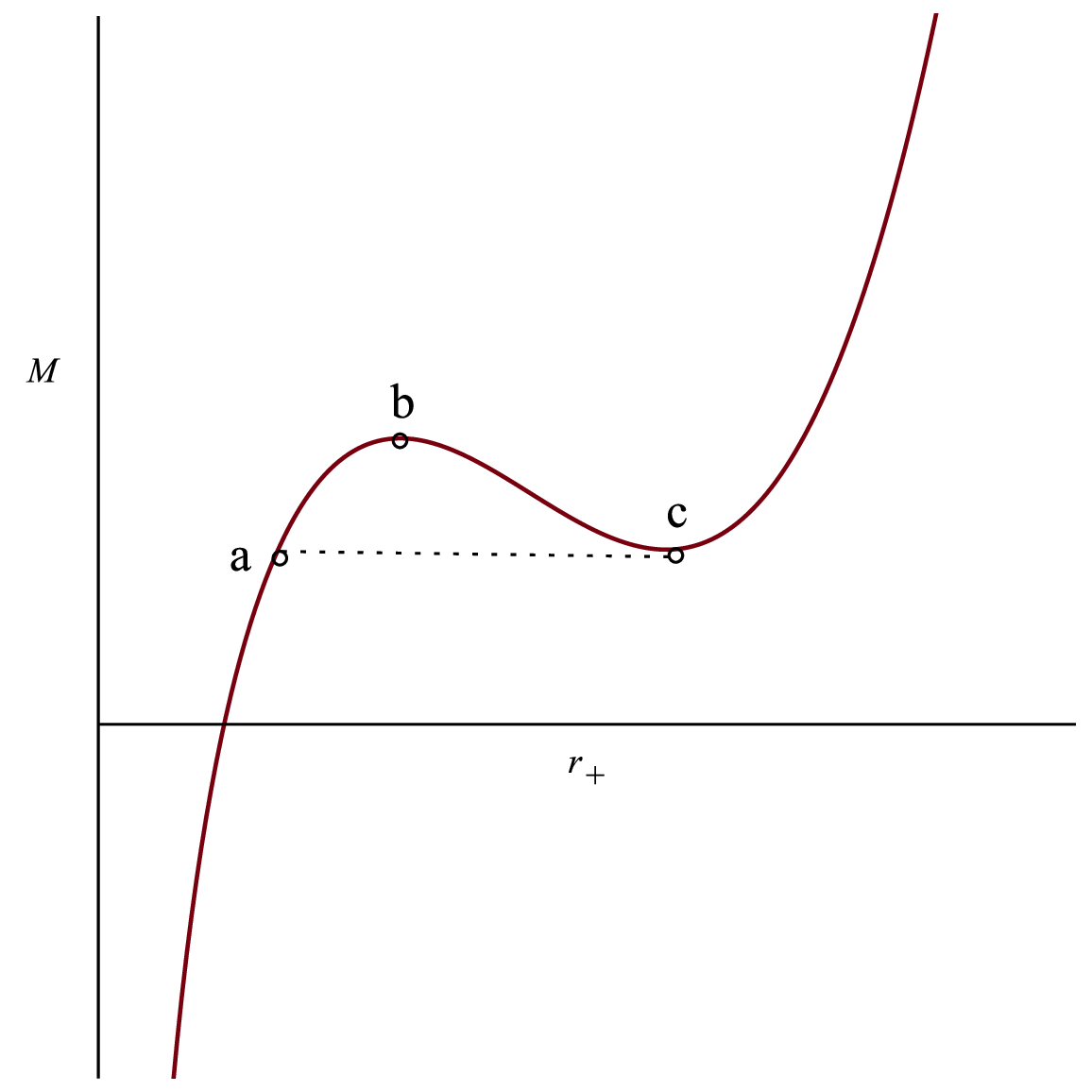}
		\vspace{-1mm}
		\caption{Schematic diagram of $M$ as $\alpha<0$ in $\gamma^2\ell^2>3(\varepsilon+1)$, $\varepsilon+1>0$, $\gamma<0$.}
		\label{fig7}
	\end{center}
\end{figure}

Next we consider the case of $\alpha<0$. Similar to the previous subsection, we give a schematic in Figure.\ref{fig7}, and we are facing the same question: where is the position of $r_+=\sqrt{-2\alpha}$? Some of our analysis in the previous section still holds. By Eq.\eqref{partial}, we can know that as long as $|\alpha|$ is large enough, we will have $\sqrt{-2\alpha}>r_c$, and the final state of the black hole is $T = \infty$ and the black hole will evaporate completely. When we decrease $|\alpha|$, the $\sqrt{-2\alpha}$ can be smaller than $r_c$. However, the most significant difference from the previous subsection is that the branch of point b to point a always exist even as $\alpha= 0$. The $r_a$ behaves as $r_a\sim|\alpha|$ when the $|\alpha|$ approaches zero in the previous subsection, while in this case the $r_a$ should reduce to a finite value, so if the $|\alpha|$ is small enough, we can have $\sqrt{-2\alpha}<r_a$, and the black hole has a new branch of solutions, i.e., from point a to $r_+=\sqrt{-2\alpha}$. 

\begin{figure}[h!]
	\begin{center}
		\includegraphics[width=0.4\textwidth]{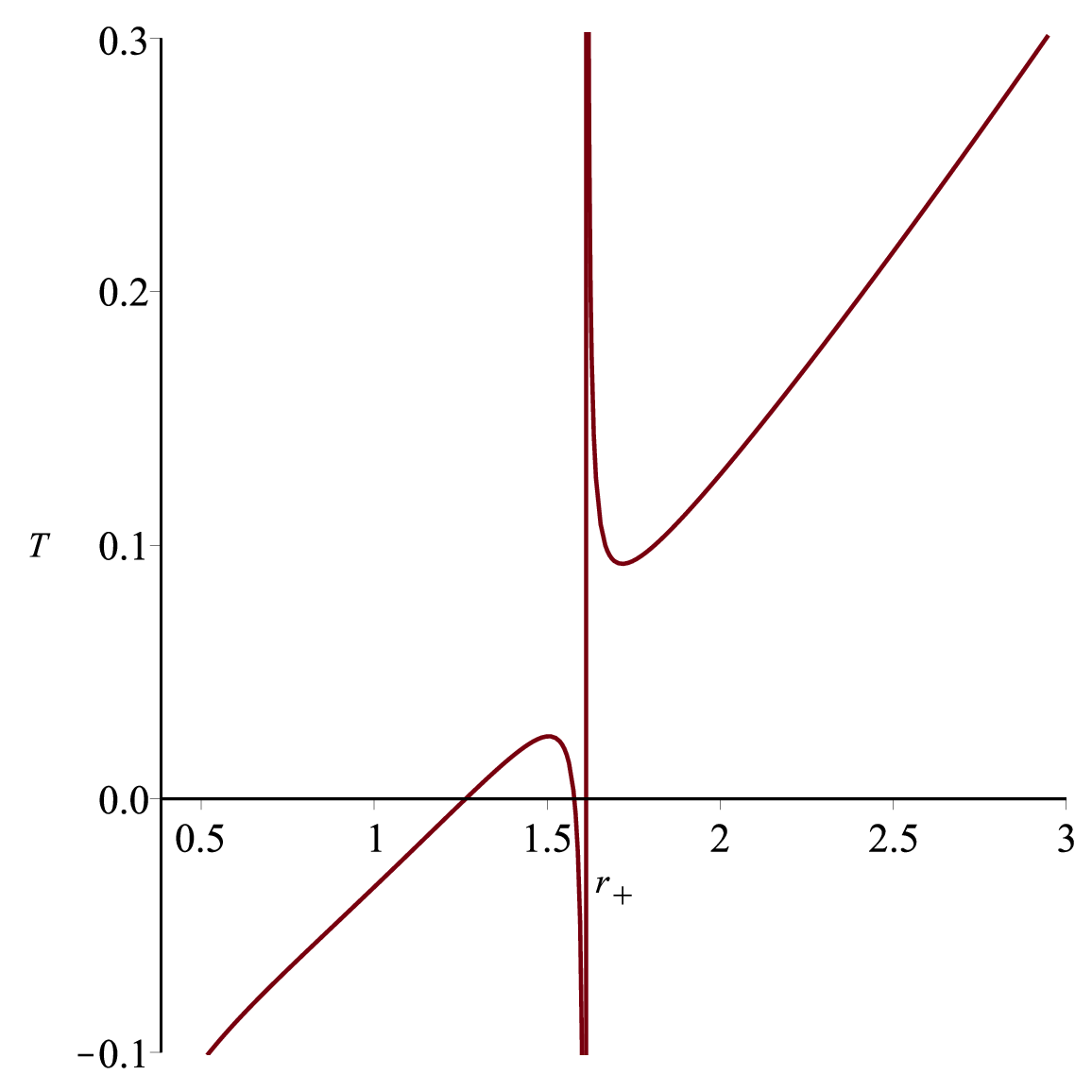}
		\includegraphics[width=0.4\textwidth]{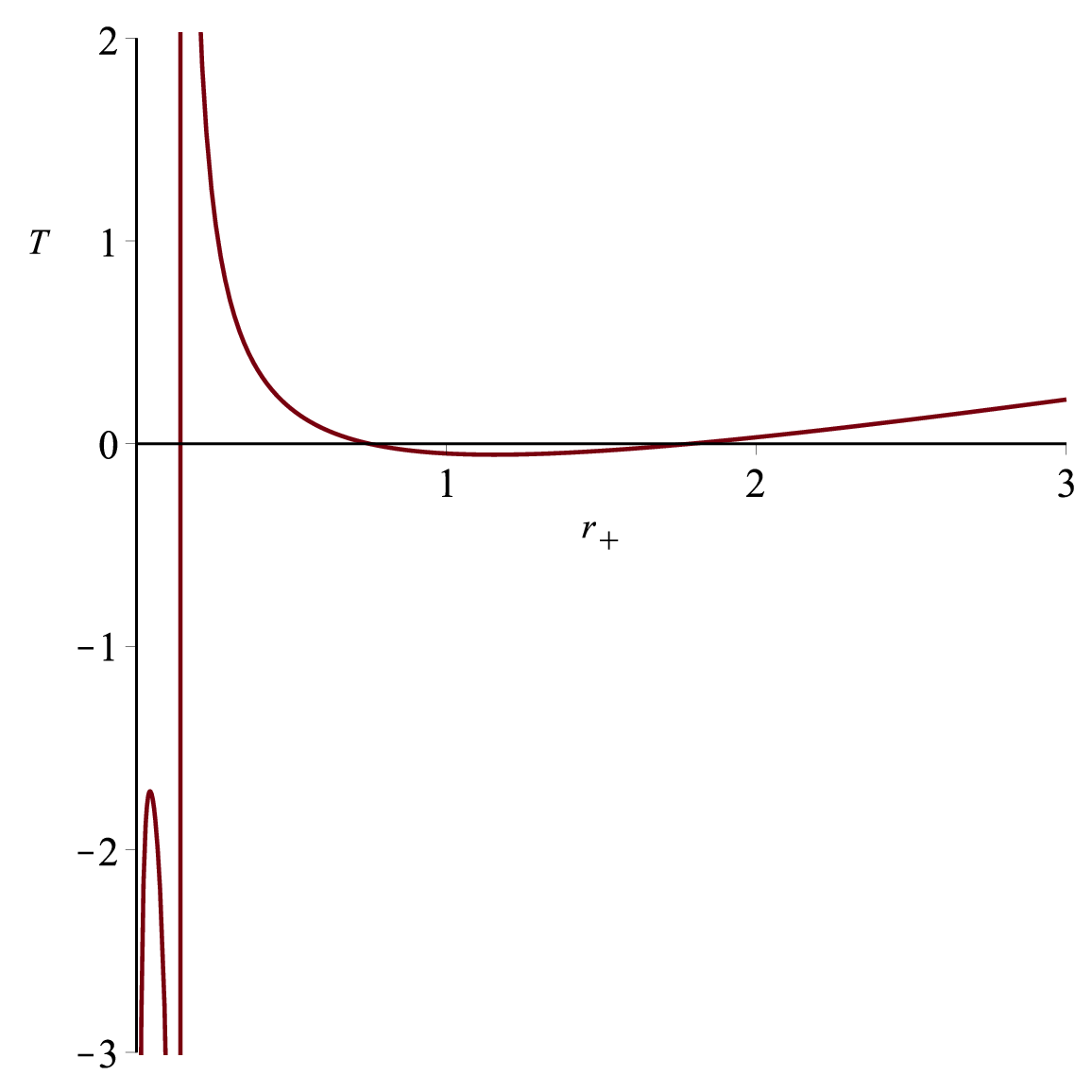}
		\vspace{-1mm}
		\caption{The black hole temperature $T$ as function of the radius $r_+$. In the first figure we set $\gamma=-3.8$, $\ell=1$, $\varepsilon=3$, $\alpha=-1.3$. In the second figure we set $\gamma=-3.8$, $\ell=1$, $\varepsilon=3$, $\alpha=-0.01$. They correspond to $\sqrt{-2\alpha}>r_c$ and $\sqrt{-2\alpha}<r_a$ respectively.}
		\label{fig8}
	\end{center}
\end{figure}

In Figure.\ref{fig8} we present two examples of the black hole temperature $T$ as the function of the radius $r_+$, corresponding to $\sqrt{-2\alpha}>r_c$ and $\sqrt{-2\alpha}<r_a$ respectively. In the first case the final state is $\sqrt{-2\alpha}$, and the black hole can evaporate completely. In the second case there are two branches of black hole solution. One is from $\infty$ to $r_c$, and the other is from $r_a$ to $\sqrt{-2\alpha}$. The first one cannot be completely evaporated, while the second one can. We will not consider the cases of $r_b<\sqrt{-2\alpha}<r_c$ and $r_a<\sqrt{-2\alpha}<r_b$ here, since they do not give any new properties.

In addition, under some parameter intervals ($\gamma \ell<-2\sqrt{\varepsilon+1}$), we will have negative minimal values of $M$ when $\alpha=0$. When $\alpha > 0$, we naturally do not have the case of the left figure in Figure. \ref{fig6}, and when $\alpha < 0$, since the point a in Figure.\ref{fig7} also no longer exists, we will not have the case of $\sqrt{-2\alpha}<r_a$. The black hole can not evaporate completely at any $\alpha$. We will not go into details here.

\section{Summary}\label{iiii}

{Using the Stefan-Boltzmann law, we can calculate the particle emission power and estimate the lifetime of the black hole. However, when gravity models become complex, the thermodynamic quantities are often accompanied by multiple parameters, and our calculations may become difficult due to the increasing degree of equations. In this work we aim to provide a general and model-independent description of the evolution of AdS black holes, focusing on the Gauss-Bonnet massive gravity. Our aim is to provide some insights and methods that can be applied to many models. Gauss-Bonnet massive gravity is a perfect example because it is a combination of two well-known models that are easy to solve independently but more difficult when combined.}

{We divide the problem into two parts. First, we investigate whether an infinitely large AdS black hole can evaporate an infinite amount of mass in finite time. Second, we study whether there are zero-temperature states that ultimately prevent evaporation. Regarding the first question, our analysis shows that the impact factor of an infinitely large AdS black hole is $b_c=\ell_{\text{eff}}$ and we prove that an infinitely large AdS black hole is always capable of evaporating an infinite amount of mass in finite time. For the second question, we adopt a strategy that avoids solving the quadratic equation directly. Instead, our strategy involves building upon our previous comprehensive analysis of massive gravity, using it as a foundation for studying the Gauss-Bonnet corrections.}

{According to the results of massive gravity, there are three cases of zero points of its temperature: zero, one, and two points. On this basis, when we introduce the Gauss-Bonnet parameter $\alpha$, it must change the properties of thermodynamic quantities. In order to classify the $T = 0$ case, we analyze the $\alpha$ corrections to the black hole mass $M$. The result shows that the sign of $\alpha$ significantly changes the black hole evaporation results. Moreover, in one of the cases (the case of 4.2), the system has an extra evaporation path for certain parameters.}

{The approach described above can also be applied to a wider range of models. Since we already know that an infinitely large AdS black hole can evaporate an infinite amount of mass in finite time, we then address the evaporation of other AdS black holes by going directly to the question of whether there are zero temperature states.  If such states exist, the black hole will become a remnant according to the third law of thermodynamics, and conversely, the black hole will evaporate with a lifetime related to the (effective) cosmological radius. If $T=0$ is a higher degree equation, we can find one model that has been rigorously solved, and use the other parameters as corrections to obtain qualitative results.} Of course, we must also emphasize that our discussion is carried out under modified gravity. If the gravity model does not include the Einstein-Hilbert action, such as in the case of conformal (Weyl) gravity \cite{Xu:2018liy}, the thermodynamics of the black hole may be very different, and we still need to consider it according to the specific thermodynamic quantities.

Finally, we emphasize that our study on black hole evaporation is still based on quantum field theory in curved spacetime and remains mainly qualitative, while solving the information loss paradox may require a deeper understanding of quantum gravity. If we want to know more about the evaporation process, we need to go into more detailed modeling and study quantum systems that can be used to describe black holes. A recent example is the Maldacena's model \cite{Maldacena:2023acv}, which is a collection of interacting oscillators and Majorana fermions. The next feasible work could be to explore the possible mass and information loss of the relevant quantum systems, so that we can simulate the evaporation of a black hole. Through numerical simulations, we may even study the information loss paradox quantitatively in the framework of open quantum systems \cite{Breuer}, and consider the interaction between the black hole and specific detectors \cite{Hodgkinson:2014iua,Ng:2018drz,Xu:2023tdt}. We hope to consider this case in our future work.

\section*{Acknowledgements}

Hao Xu thanks National Natural Science Foundation of China (No.12205250) for funding support.

\noindent  \textbf{Data Availability Statement}: No Data associated in the manuscript.


\providecommand{\href}[2]{#2}\begingroup
\footnotesize\itemsep=0pt
\providecommand{\eprint}[2][]{\href{http://arxiv.org/abs/#2}{arXiv:#2}}

\end{document}